\documentclass[11pt,a4paper]{article}

\usepackage[T1]{fontenc}
\usepackage[utf8]{inputenc}
\usepackage{lmodern}
\usepackage{amsmath,amssymb,amsfonts}
\usepackage{graphicx}
\graphicspath{{figs/}}
\usepackage{booktabs}
\usepackage{multirow}
\usepackage{xspace}
\usepackage[margin=2.5cm]{geometry}
\usepackage[backend=biber,style=numeric-comp,sorting=none,giveninits=true,
  maxbibnames=8,minbibnames=6,doi=true,eprint=true,url=true,isbn=false,
  datamodel=collab]{biblatex}
\renewbibmacro{in:}{}
\renewbibmacro*{begentry}{\iffieldundef{collaboration}{}{\printfield{collaboration}\space Collaboration,\space}}
\DeclareFieldFormat{eprint:arxiv}{arXiv:\href{https://arxiv.org/abs/#1}{#1}\iffieldundef{eprintclass}{}{\space[\thefield{eprintclass}]}}
\usepackage[colorlinks=true,linkcolor=blue,citecolor=blue,urlcolor=blue]{hyperref}
\usepackage{nccmath}
\usepackage{xcolor}
\usepackage{tikz}
\usetikzlibrary{decorations.pathmorphing,decorations.markings}
\tikzset{
  fermion/.style={draw, postaction={decorate},
    decoration={markings, mark=at position 0.55 with {\arrow{stealth}}}},
  gluon/.style={draw, decorate,
    decoration={coil, aspect=0.6, segment length=4pt, amplitude=2.5pt}},
  vboson/.style={draw, decorate,
    decoration={snake, segment length=6pt, amplitude=1.6pt}},
  vtx/.style={circle, fill=black, inner sep=1pt},
}

\allowdisplaybreaks

\newcommand{\dyturbo}{\textsc{DYTurbo}\xspace}
\newcommand{\mcfm}{\textsc{MCFM}\xspace}
\newcommand{\form}{\textsc{Form}\xspace}
\newcommand{\kira}{\textsc{Kira}\xspace}
\newcommand{\firefly}{\textsc{FireFly}\xspace}
\newcommand{\ratracer}{\textsc{Ratracer}\xspace}
\newcommand{\sympy}{\textsc{SymPy}\xspace}
\newcommand{\qT}{q_{\mathrm{T}}}
\newcommand{\qTcut}{q_{\mathrm{T}}^{\mathrm{cut}}}
\newcommand{\pT}{p_{\mathrm{T}}}
\newcommand{\as}{\alpha_{\mathrm{s}}}
\newcommand{\muR}{\mu_{\mathrm{R}}}
\newcommand{\muF}{\mu_{\mathrm{F}}}
\newcommand{\sigUL}{\sigma^{U+L}}
\newcommand{\stw}{\sin^2\theta_{\mathrm{W}}}
\newcommand{\mW}{m_W}
\newcommand{\CS}{Collins--Soper\xspace}
\newcommand{\order}[1]{\mathcal{O}\!\left(#1\right)}

\title{\bf Analytic next-to-leading-order helicity cross sections\\
for vector-boson production at finite transverse momentum}

\newcommand{\orcidid}[1]{\textsuperscript{\href{https://orcid.org/#1}{\,\textcolor{teal}{iD}}}}

\date{}

\begin{document}
\begin{center}
  {\Large\bfseries Analytic next-to-leading-order helicity cross sections\\[1mm]
   for vector-boson production at finite transverse momentum\par}
  \vspace{7mm}
  {\bfseries Stefano Camarda}${}^{(a)}$\orcidid{0000-0003-0479-7689},
  {\bfseries Leandro Cieri}${}^{(b)}$\orcidid{0000-0002-2624-1879},\\[1mm]
  {\bfseries Giancarlo Ferrera}${}^{(c)}$\orcidid{0000-0002-4559-0740}
  and {\bfseries Lorenzo Rossi}${}^{(c,d)}$\orcidid{0000-0002-8326-3118}\par
  \vspace{5mm}
  \begin{minipage}{0.92\textwidth}\centering\small
    ${}^{(a)}$ CERN, CH-1211 Geneva, Switzerland\\[1.5mm]
    ${}^{(b)}$ Instituto de F\'isica Corpuscular, Universitat de Val\`encia --
    Consejo Superior de Investigaciones Cient\'ificas,
    Parc Cient\'ific, E-46980 Paterna, Valencia, Spain\\[1.5mm]
    ${}^{(c)}$ Dipartimento di Fisica, Universit\`a di Milano and INFN,
    Sezione di Milano, I-20133 Milan, Italy\\[1.5mm]
    ${}^{(d)}$ Institute for Theoretical Physics, Universit\"at T\"ubingen,
    Auf der Morgenstelle 14, D-72076 T\"ubingen, Germany
  \end{minipage}
  \vspace{4mm}
\end{center}

\begin{abstract}
\noindent
We present a complete analytic calculation of the next-to-leading-order
(NLO) QCD cross sections for the production of an electroweak vector
boson ($W^\pm$, $Z/\gamma^*$) at finite transverse momentum, decomposed
into the full set of helicity cross sections that determine the angular
coefficients $A_0$--$A_7$ of the lepton-pair angular distribution in the
\CS frame. The calculation includes the unpolarised
cross section and all eight helicity projections for both
neutral- and charged-current Drell--Yan production, the axial
quark-triangle contributions in the $m_t\to\infty$ limit and with full
top-quark-mass dependence, and closed-form expressions for all
coefficient functions. The results are implemented in the \dyturbo
program and validated at the level of a few times $10^{-4}$ against its numerical
implementation derived from \mcfm, as well as through the analytic
$\qT\to0$ limit. The analytic implementation achieves a speed-up of
more than five orders of magnitude with respect to the numerical
code,
enabling fast and numerically stable predictions for precision
electroweak measurements at the LHC. The calculation was carried out
with extensive use of large language models (Anthropic Claude) as
autonomous agents; we document the methodology and the associated
computational and token budgets.
\end{abstract}

\tableofcontents
\clearpage

\section{Introduction}
\label{sec:intro}

The Drell--Yan~\cite{Drell:1970wh,Christenson:1970um} production of electroweak vector bosons,
$pp \to W^\pm/Z/\gamma^* + X \to \ell\ell' + X$ --- where $\ell\ell'$
denotes a pair of charged leptons for the neutral current and a
charged lepton and a neutrino for the charged current --- is a
cornerstone of the
physics programme of the Large Hadron Collider (LHC). Its large cross
section, clean experimental signature, and the availability of
higher-order QCD and electroweak corrections make it the primary tool
for precision measurements of
fundamental Standard Model parameters, most notably the $W$-boson mass
$\mW$~\cite{D0:2013jba,ATLAS:2017rzl,ATLAS:2024erm,CDF:2022hxs,CMS:2024lrd,LHCb:2021bjt,LHC-TeVMWWorkingGroup:2023zkn}
and the weak mixing angle
$\stw$~\cite{CDF:2018cnj,ATLAS:2018gqq,CMS:2024ony,LHCb:2024ygc}. The precision
targets of these measurements --- $\order{10~\mathrm{MeV}}$ on $\mW$ and
$\order{10^{-4}}$ on $\stw$ --- place extreme demands on the
theoretical description of the production dynamics, and in particular
on the modelling of the vector-boson transverse momentum $\qT$ and of
the angular distribution of the decay leptons.

The full five-fold differential cross section factorises into a boson
production part and a decay part expressed through nine helicity cross
sections. In the rest frame of the lepton pair, the decay-angle
dependence is captured by
the angular coefficients $A_0$--$A_7$, ratios of helicity cross
sections to the unpolarised ($U{+}L$) cross section; throughout this
paper the decay angles are defined in the \CS
frame~\cite{Collins:1977iv}. The angular
coefficients directly control the lepton kinematics used in precision
fits: the lepton $\pT$ spectrum in $\mW$ measurements, and the
forward--backward asymmetry underlying $\stw$ determinations through
$A_4$. They have been measured in fixed-target experiments by
NA10~\cite{NA10:1986fgk} and E615~\cite{E615:1989bda}, at the Tevatron by
CDF~\cite{CDF:2005qwt,CDF:2011ksg} and with high precision at the LHC
by ATLAS~\cite{ATLAS:2016rnf,ATLAS:2023lsr,ATLAS:2025wai},
CMS~\cite{CMS:2015cyj,CMS:2026amb} and LHCb~\cite{LHCb:2022tbc}. The LHC
measurements revealed, in particular, a significant
violation of the Lam--Tung relation~\cite{Lam:1978pu,Lam:1978zr,Lam:1980uc}
($A_0 \neq A_2$) that is a genuine
$\order{\as^2}$ effect~\cite{Lambertsen:2016wgj}.

At finite $\qT$, the helicity cross sections start at $\order{\as}$
through the $V{+}$jet process. The leading-order (LO) structure
functions were computed long
ago~\cite{Collins:1977iv,Lam:1978pu,Lam:1978zr}, and
the next-to-leading-order (NLO) QCD corrections, of $\order{\as^2}$,
were computed analytically for the unpolarised cross section, first
for a non-singlet combination in virtual-photon production by Ellis,
Martinelli and Petronzio~\cite{Ellis:1981hk} and then in full in the
seminal works of Gonsalves, Pawlowski and Wai~\cite{Gonsalves:1989ar}
and of Arnold and Reno~\cite{Arnold:1988dp}; for the angular
coefficients by Mirkes and
Ohnemus~\cite{Mirkes:1990diss,Mirkes:1992hu,Mirkes:1994dp,Mirkes:1994eb};
and for the T-odd coefficients by Hagiwara, Hikasa and
Kai~\cite{Hagiwara:1984hi} (recently
revisited in Ref.~\cite{Lyubovitskij:2024civ}) ---
calculations of the same analytic kind as
the present one. Numerical Monte Carlo implementations followed, in
\mcfm~\cite{Campbell:1999ah,Campbell:2002tg} and
NNLOJET~\cite{Gauld:2017tww}, together with NNLO
calculations of $V{+}$jet
production~\cite{Ridder:2015dxa,Boughezal:2015dva,Boughezal:2015ded,Pellen:2022fom},
which in \mcfm also underlie the N$^3$LO Drell--Yan predictions of
Refs.~\cite{Neumann:2022lft,Campbell:2023lcy}.
The low-$\qT$ region of the angular coefficients, where the
fixed-order description must be supplemented by transverse-momentum
resummation~\cite{Dokshitzer:1978yd,Parisi:1979se,Collins:1981uk,Collins:1984kg}
and transverse-momentum-dependent factorisation, and the
small-$\qT$ power expansion of the fixed-order result itself, have been
studied in
Refs.~\cite{Boer:2006eq,Berger:2007si,Berger:2007jw,Arnold:2008kf,Bacchetta:2008xw,Peng:2015spa,Chang:2017kuv,Chang:2018pvk,Peng:2018tty,Peng:2019boj,Bacchetta:2019qkv,Ebert:2021jhy,Vladimirov:2023aot,Lyubovitskij:2024jlb,Piloneta:2024aac,Lyubovitskij:2025oig,Camarda:2026vjq,Gramotkov:2026sci}.
The numerical implementations rely on Monte Carlo
integration over the real-emission phase space, with residual
statistical noise that is strongly amplified when forming the angular
coefficients --- ratios of small helicity differences over the
unpolarised cross section. For fitting applications, in which the predictions must be
evaluated at many points in parameter space with sub-permille numerical
stability, this cost is prohibitive.

In this paper we present a complete analytic calculation of the
NLO QCD helicity cross sections for $W^\pm$ and $Z/\gamma^*$ production
at finite $\qT$: the unpolarised cross section
$\sigUL$ and all eight helicity projections
$\sigma_0,\ldots,\sigma_7$, for both neutral-current and
charged-current Drell--Yan. All partonic coefficient functions are
obtained in closed form; the remaining convolutions with the parton
distribution functions are performed with deterministic quadrature
rules, yielding predictions free of Monte Carlo noise, with an
evaluation cost reduced by more than five orders of magnitude with
respect to the numerical code. The calculation includes the parity-violating and
naive-time-reversal-odd coefficients $A_5$--$A_7$, which first appear
at $\order{\as^2}$ through absorptive parts of the one-loop
amplitudes~\cite{Hagiwara:1984hi}, and the axial quark-triangle
contributions, both in the $m_t\to\infty$ limit and with exact
top-mass dependence.

Several of these results are genuinely new. Numerical
$\order{\as^2}$ predictions for the parity-odd coefficients $A_3$,
$A_4$ were shown in Ref.~\cite{Mirkes:1994dp}, but no analytic
expressions for their coefficient functions ever appeared in print,
and part of the four-quark sector appears only in a doctoral
thesis~\cite{Mirkes:1990diss}, which we obtained in the course of
this work and use as an analytic cross-check. The axial
quark-triangle contributions
were previously available in closed form only for the unpolarised
cross section and only in the $m_t\to\infty$
limit~\cite{Gonsalves:1989ar}, while the exact massive-top one-loop
amplitudes for $Z{+}$jet production were computed in
Ref.~\cite{Campbell:2016tcu}; here the triangle contributions are
given in closed form for all helicity projections and with exact
top-mass dependence. And --- most importantly for practical use --- no public
code implementing the analytic NLO helicity cross sections was ever
released; the present implementation in \dyturbo is the first.

The results are implemented in the \dyturbo
program~\cite{Camarda:2019zyx}, alongside its established numerical
$V{+}$jet implementation inherited from
\mcfm~\cite{Campbell:1999ah}, which provides an independent
validation reference: we demonstrate agreement at the level of a
few times $10^{-4}$ across the kinematic range of the LHC
measurements,
for all helicity cross sections and both boson species. A second,
internal validation exploits the analytically known $\qT\to0$
singular behaviour: the helicity cross sections must match the asymptotic term of
transverse-momentum resummation --- the fixed-order
expansion of the resummed cross section, for which we use the
Catani--de Florian--Grazzini
formalism~\cite{Catani:2000vq,Bozzi:2005wk} --- for
$\sigUL$ and $\sigma_4$ ($A_4$), and remain finite or vanish for the
other projections. This limit is a powerful diagnostic of the entire
calculation chain.

A distinctive aspect of this work is its methodology: the calculation
was carried out with extensive, systematic use of large language
models (LLMs) of the Anthropic Claude family (Claude Fable, Claude
Opus, and Claude
Sonnet)~\cite{Anthropic:2026fable5,Anthropic:2026fable51,Anthropic:2026opus48,Anthropic:2026opus5,Anthropic:2026opus55,Anthropic:2026sonnet5}
operating as autonomous agents in the Claude Code
environment~\cite{Anthropic:2026code}. The agents performed the symbolic derivations
(driving \form~\cite{Vermaseren:2000nd,Tentyukov:2007mu,Kuipers:2012rf,Ueda:2020wqk,Davies:2026cci},
\kira~\cite{Maierhofer:2017gsa,Klappert:2020nbg,Lange:2025fba}, \firefly~\cite{Klappert:2019emp,Klappert:2020aqs},
\ratracer~\cite{Magerya:2022hvj}, and
\sympy~\cite{Meurer:2017yhf} computations), generated and verified
the C++ implementation, designed and ran the validation campaigns, and
maintained the project documentation across several hundred working
sessions. Every analytic result was subjected to independent
numerical verification before being accepted; the human authors defined the
physics goals, arbitrated the validation criteria, and reviewed the
results. We report the methodology and the associated computational
and token budgets as a data point on the use of LLM agents in
perturbative QCD calculations. This adds to a rapidly growing body of
work in which LLMs have contributed to physics results: the proof of
an identity between the critical exponents of
jamming~\cite{Parisi:2026jam}, the discovery that single-minus gluon
tree amplitudes are nonvanishing~\cite{Guevara:2026qzd},
differential reduction of Feynman integrals~\cite{Britto:2026qjn},
the NLL+NLO resummation of the C-parameter Sudakov shoulder, carried
out entirely by an LLM under supervision~\cite{Schwartz:2026ekw}, and,
in experimental particle physics, autonomous analyses of open
collider data by LLM agents~\cite{Moreno:2026mqk}.

The paper is organised as follows. Section~\ref{sec:formalism}
introduces the helicity decomposition, the \CS frame, and the
structure of the NLO calculation. Section~\ref{sec:methodology}
describes the computational methodology and the software employed.
Section~\ref{sec:results} presents the validation against the
numerical \dyturbo/\mcfm implementation.
Section~\ref{sec:triangles} discusses the axial triangle
contributions. Section~\ref{sec:lowqt} presents the $\qT\to0$ limit
and the cancellation against the asymptotic term.
Section~\ref{sec:costs} documents
CPU and LLM-token budgets, and Section~\ref{sec:conclusions} contains
our conclusions. Appendix~\ref{app:notation} collects the notation and the
framework formulas (kinematics, projectors, couplings and
luminosities, phase space, and master integrals);
Appendix~\ref{app:coefficients} defines the notation of the
closed-form coefficient functions, which are provided as ancillary
files;
Appendix~\ref{app:misprints} records misprints found in the
literature.

\section{Helicity cross sections and angular coefficients}
\label{sec:formalism}

\subsection{Angular decomposition in the Collins--Soper frame}
\label{sec:formalism:angular}

We consider the production of a lepton pair through an intermediate
electroweak boson,
\begin{equation}
  h_1(P_1) + h_2(P_2) \;\to\; V(q) + X \;\to\; \ell(l_1)\,\bar\ell'(l_2) + X,
  \qquad V = W^\pm,\, Z/\gamma^*,
\label{eq:process}
\end{equation}
at finite transverse momentum $\qT$ of the boson, with virtuality
$q^2$ and rapidity $y$; we write $m = \sqrt{q^2}$ for the
invariant mass of the lepton pair. Each helicity cross section is computed
in QCD-improved parton-model factorisation,
\begin{equation}
  \frac{\mathrm{d}\sigma_i}{\mathrm{d}q^2\,\mathrm{d}\qT\,\mathrm{d}y}
  = \sum_{a,b} \int_0^1 \mathrm{d}x_1\,\mathrm{d}x_2\;
    f_a^{h_1}\!(x_1,\muF)\, f_b^{h_2}\!(x_2,\muF)\;
    \mathrm{d}\hat\sigma_{i,ab}
    \bigl(x_1 P_1, x_2 P_2;\, \as(\muR), \muF\bigr) ,
\label{eq:factorization}
\end{equation}
where $P_{1,2}$ are the hadron momenta, $x_{1,2}$ the momentum
fractions of the partons in the hadrons, $a,b$ run over quarks,
antiquarks and gluons, $f_a^{h}(x,\muF)$ are the parton distribution
functions, $\muF$ and $\muR$ the factorisation and renormalisation
scales, and $\as$ the strong coupling; $\ell(l_1)$ denotes the decay
fermion --- the negatively charged lepton for $Z/\gamma^*$ and
$W^-$, the neutrino for $W^+$ --- and $\bar\ell'(l_2)$ the
accompanying antifermion. The partonic
cross sections $\mathrm{d}\hat\sigma_{i,ab}$ are the objects computed
analytically in this paper. Kinematic notation (Mandelstam invariants
at hadron and parton level and
the relations fixing them from $(q^2,\qT,y)$ and the momentum
fractions) is collected in Appendix~\ref{app:notation:kinematics}.

To all orders in QCD the
five-fold differential cross section factorises into nine
helicity cross sections multiplying the corresponding spherical
harmonics of the decay angles. Expressed in terms of the polar and azimuthal
angles $(\theta,\phi)$ of the lepton $\ell$ of
Eq.~\eqref{eq:process} in the \CS
rest frame of the pair~\cite{Collins:1977iv}, it reads
\begin{align}
\frac{\mathrm{d}\sigma}{\mathrm{d} q^2\,\mathrm{d} \qT\,\mathrm{d} y\,\mathrm{d}\cos\theta\,\mathrm{d}\phi}
={}& \frac{3}{16\pi}
\Big[ \sigUL (1+\cos^2\theta)
   + \sigma_0\, \tfrac{1}{2} (1-3\cos^2\theta)
   + \sigma_1 \sin 2\theta \cos\phi \nonumber\\
&  + \sigma_2\, \tfrac{1}{2} \sin^2\theta \cos 2\phi
   + \sigma_3 \sin\theta \cos\phi
   + \sigma_4 \cos\theta
   + \sigma_5 \sin^2\theta \sin 2\phi \nonumber\\
&  + \sigma_6 \sin 2\theta \sin\phi
   + \sigma_7 \sin\theta \sin\phi \Big],
\label{eq:angulardecomposition}
\end{align}
where the $\sigma_i \equiv
\mathrm{d}\sigma_i/(\mathrm{d} q^2\,\mathrm{d}\qT\,\mathrm{d} y)$ are the unnormalised
helicity cross sections and the angular coefficients are their
ratios to the unpolarised cross section,
\begin{equation}
  A_i = \frac{\sigma_i}{\sigUL} .
\end{equation}
Throughout, the $A_i$ are formed as unexpanded ratios of the
helicity cross sections evaluated at the same order in $\as$
(both through $\order{\as^2}$ at NLO); for $A_5$--$A_7$, whose
numerator starts at $\order{\as^2}$, the prediction is thus of
leading order in the coefficient while the denominator is the NLO
one.
All harmonics with $i \ge 0$ integrate to zero over the full solid
angle, so integrating Eq.~\eqref{eq:angulardecomposition} over
$(\cos\theta,\phi)$ returns $\sigUL$ exactly; equivalently the
$A_i$ are weighted moments of the decay distribution, e.g.\
$A_4 = 4\langle\cos\theta\rangle$,
$A_3 = 4\langle\sin\theta\cos\phi\rangle$,
$A_2 = 10\langle\sin^2\theta\cos2\phi\rangle$.

The angles are those of the lepton $\ell$ of
Eq.~\eqref{eq:process} --- for $W^+$ production, the neutrino ---
and the frame
conventions are implemented as follows: the \CS polar axis bisects
the two beam directions in the dilepton rest frame, with $\cos\theta$
carrying a factor $\mathrm{sign}(p_z^V)$, with $p_z^V$ the boson
longitudinal momentum, so that the polar axis
follows the boson longitudinal boost; the azimuthal $x$-axis lies
along the unit vector $\hat q_{\mathrm{T}}$ of the boson transverse
momentum in the event plane,
and the $y$-axis along $\hat z\times\hat q_{\mathrm{T}}$, with $\hat z$
the beam axis, carrying the same
$\mathrm{sign}(p_z^V)$ factor so that the triad is right-handed for
either sign of the rapidity. For $A_0$--$A_4$ these are the
conventions of the LHC
measurements~\cite{ATLAS:2016rnf,CMS:2015cyj,ATLAS:2025wai}; in
particular, Ref.~\cite{ATLAS:2025wai} uses the neutrino direction
for $W^+$. For $A_5$--$A_7$ we quote the sign that follows from this
triad, which is opposite to the one used by the LHC measurements:
the two choices differ by the orientation of the azimuthal $y$-axis,
that is by $\phi\to-\phi$, which reverses exactly the $\sin\phi$
and $\sin2\phi$ harmonics and leaves the $\cos$ harmonics
invariant. Note that for $W^+$ this differs from the convention of
Ref.~\cite{Mirkes:1992hu}, which uses the charged lepton; the two
choices are related by $(\theta,\phi)\to(\pi-\theta,\phi+\pi)$,
which reverses the sign of $A_3$, $A_4$ and $A_7$. The
coefficients $A_0$--$A_4$ are even under naive time reversal and
start at $\order{\as}$. The coefficients
$A_5$--$A_7$ are T-odd --- odd under naive time reversal, which
reverses momenta and spins without exchanging initial and final
states --- and first arise at $\order{\as^2}$ from the
absorptive parts of the one-loop amplitudes~\cite{Hagiwara:1984hi}.
We refer to $A_0$--$A_4$ as the T-even sector, subdivided into the
parity-even $A_0$--$A_2$ and the parity-odd $A_3$, $A_4$, and to
$A_5$--$A_7$ as the T-odd sector.
At $\order{\as}$
the Lam--Tung relation~\cite{Lam:1978pu} $A_0 = A_2$ holds; its
violation is generated at $\order{\as^2}$.

\subsection{Projectors and hadronic structure functions}
\label{sec:formalism:projectors}

The helicity cross sections are obtained from the hadronic tensor
$H^{\mu\nu}$ by covariant projection. For the parity-conserving
sector ($\sigUL$, $A_0$, $A_1$, $A_2$) we follow the projector basis of
Ref.~\cite{Mirkes:1992hu}: the four raw contractions
\begin{equation}
  I_{U+L} = -g_{\mu\nu} H^{\mu\nu}, \qquad
  I_{L1} = p_{1\mu} p_{1\nu} H^{\mu\nu}, \qquad
  I_{L2} = p_{2\mu} p_{2\nu} H^{\mu\nu}, \qquad
  I_{L12} = \bigl(p_{1\mu} p_{2\nu} + p_{2\mu} p_{1\nu}\bigr) H^{\mu\nu},
\label{eq:evenprojectors}
\end{equation}
with $g_{\mu\nu}$ the metric tensor and $p_{1,2}$ the incoming
parton momenta, and with the terms proportional to $q^\mu$ dropped by virtue of
current conservation (verified as exact Ward identities per
channel), are normalised and rotated to the \CS frame.
The parity-odd sector ($A_3$, $A_4$) is obtained from the
antisymmetric part of the tensor through the Levi-Civita
contractions
\begin{equation}
  I_{P1} = \varepsilon_{\mu\nu\alpha\beta}\, q^\alpha p_1^\beta\, H^{\mu\nu},
  \qquad
  I_{P2} = \varepsilon_{\mu\nu\alpha\beta}\, q^\alpha p_2^\beta\, H^{\mu\nu},
\label{eq:oddprojectors}
\end{equation}
and the T-odd sector ($A_5$, $A_6$, $A_7$) through the wedge
contractions built from the out-of-plane vector
$W^\mu = \varepsilon^{\mu\nu\rho\sigma} q_\nu p_{1\rho} p_{2\sigma}$,
with $\varepsilon^{\mu\nu\rho\sigma}$ the totally antisymmetric
tensor,
\begin{equation}
  I_{W1} = W_\mu p_{1\nu}\, H^{\mu\nu}, \qquad
  I_{W2} = W_\mu p_{2\nu}\, H^{\mu\nu}, \qquad
  I_{W3} = \bigl(p_{1\mu} p_{2\nu} - p_{2\mu} p_{1\nu}\bigr) H^{\mu\nu},
\label{eq:toddprojectors}
\end{equation}
evaluated with an anticommuting $\gamma_5$ in
strictly four dimensions --- legitimate because the T-odd
projections are separately ultraviolet- and infrared-finite.
The parity-odd projections of Eq.~\eqref{eq:oddprojectors}, in
contrast, are infrared divergent and are evaluated in $d$
dimensions. Since, outside the axial-triangle class of
Section~\ref{sec:triangles}, $\gamma_5$ appears only on a single open
massless quark line, the real-emission contributions and the
mass-factorisation counterterm use an anticommuting $\gamma_5$,
while the one-loop virtual corrections use the Larin
prescription~\cite{Larin:1993tq} with the finite renormalisation
of the non-singlet axial current that restores the
anticommuting-$\gamma_5$ Ward identity.

The partonic helicity cross sections of Eq.~\eqref{eq:factorization}
are assembled from the projected coefficient functions as
\begin{equation}
  \frac{\mathrm{d}\sigma_i}{\mathrm{d}q^2\,\mathrm{d}\qT\,\mathrm{d}y}
  \;=\; \sum_{ab} \int\!\mathrm{d}x_1\,\mathrm{d}x_2\;
  \mathcal{L}_{ab}(x_1,x_2;\muF)\;
  \sum_{\beta} M_{i\beta}\,
  C^{\,\beta}_{ab}(s,t,u,q^2,s_2;\as(\muR),\muF) ,
\label{eq:master}
\end{equation}
in terms of the partonic Mandelstam invariants $s = (p_1+p_2)^2$,
$t = (p_1-q)^2$, $u = (p_2-q)^2$, with $p_i = x_i P_i$ --- capital
letters denoting the corresponding hadron-level invariants, so that
$s = x_1x_2 S$ with $\sqrt{S}$ the hadronic centre-of-mass energy
--- and
of the invariant mass squared of the recoiling partonic system,
\begin{equation}
  s_2 = s + t + u - q^2 ,
\end{equation}
which vanishes at LO and at the soft and collinear endpoints of the
NLO real emission, and is fixed by $x_1$, $x_2$ and the boson
kinematics. The luminosity $\mathcal{L}_{ab}$ collects the parton densities
of the channel $ab$ with its electroweak couplings and boson
propagators; the coefficient functions $C^\beta_{ab}$ are the
projections of the partonic tensor on the contractions of
Eqs.~\eqref{eq:evenprojectors}--\eqref{eq:toddprojectors}, labelled
by $\beta$ (the projection basis), carry the powers of $\as(\muR)$, the flux and
phase-space factors and are distributions in $s_2$
(Section~\ref{sec:formalism:nlo}); and $M_{i\beta}$ is the conversion
from the projection basis to the helicity cross sections. The absolute normalisation
is that of Ref.~\cite{Mirkes:1992hu}, whose $\order{\as}$ term is the
$V{+}$jet cross section of Ref.~\cite{Gonsalves:1989ar}. The
four-quark channels split into gauge-invariant classes with distinct
coupling structures, over which $C^\beta_{ab}\mathcal{L}_{ab}$ is a
sum. The kinematic relations, the normalisation of the projections,
the conversion $M_{i\beta}$, the covariant construction of the \CS
axes and the class decomposition of the four-quark channels are given
in Appendix~\ref{app:notation}.

\subsection{Structure of the NLO calculation}
\label{sec:formalism:nlo}

At finite $\qT$ the helicity cross sections start at $\order{\as}$
with the $2\to2$ channels $q\bar q \to Vg$ and $qg \to Vq$ (and
crossings). At $\order{\as^2}$ they receive the one-loop virtual
corrections to these channels and the real emission of a second
parton: $q\bar q \to Vgg$, $qg \to Vqg$, $gg \to Vq\bar q$ (in
the channel labels $qg$, $gq$, $gg$ the first letter is the parton
from hadron 1), and
the four-quark channels, organised in the annihilation and
scattering classes of
Appendix~\ref{app:notation:couplings}, including the identical-quark
interferences; the same class functions serve the neutral- and
charged-current cases through their luminosity realizations.

The real-emission phase space is integrated analytically. With the
boson kinematics $(q^2, \qT, y)$ held fixed, the recoil mass $s_2$ of
Eq.~\eqref{eq:master} is determined by the incoming momentum
fractions, so the only nontrivial integration is the collinear one,
and the coefficient functions of Eq.~\eqref{eq:master} take the
form of distributions in $s_2$,
\begin{equation}
  C^{\,\beta}_{ab}
  = C^{\,\beta,\delta}_{ab}\,\delta(s_2)
  + C^{\,\beta,1}_{ab}\left[\frac{1}{s_2}\right]_+
  + C^{\,\beta,\log}_{ab}\left[\frac{\ln(s_2/q^2)}{s_2}\right]_+
  + C^{\,\beta,\mathrm{reg}}_{ab}(s_2),
\label{eq:s2structure}
\end{equation}
where the dependence on $s$, $t$, $u$ and $q^2$ is
understood, the coefficients $C^{\beta,\delta}$,
$C^{\beta,1}$, $C^{\beta,\log}$ and $C^{\beta,\mathrm{reg}}$ are
closed-form functions of the partonic invariants (the first three
evaluated at $s_2 = 0$), per channel and helicity projection, and $[\,\cdot\,]_+$ the plus distribution
in $s_2$ over its kinematic range at fixed $x_1$, defined following
Refs.~\cite{Gonsalves:1989ar,Mirkes:1992hu} in
Appendix~\ref{app:notation:phasespace}. The
infrared poles cancel algebraically, as exact symbolic
identities: the $s_2 > 0$ collinear poles against the
PDF mass-factorisation counterterm point-wise in $s_2$, and the
endpoint $\delta(s_2)$ poles against the virtual corrections
(verified against the universal infrared singularity structure of
one-loop amplitudes~\cite{Catani:1998bh}). As a
consequence the analytic NLO cross section requires no numerical
integration beyond the PDF convolutions, which at NLO carry one
dimension more than at LO: the integration over $s_2$, which is
fixed to zero at LO. The phase-space
factorisation, the change of variables that trades $x_2$ for
$s_2$, the plus-distribution prescriptions, and the angular and
loop master integrals behind the coefficient functions are
collected in Appendix~\ref{app:notation}.

The T-odd sector is structurally special: at $\order{\as^2}$ the
coefficients $A_5$--$A_7$ are generated exclusively by the
absorptive part of the one-loop $V{+}$jet amplitude interfered
with the Born amplitude. The real-emission contribution is T-even
at this order --- T-odd real-emission contributions first arise at
$\order{\as^3}$, through the absorptive parts of the one-loop
corrections to the real emission --- so there are no
plus-distributions and no mass-factorisation counterterm, and every absorptive pole cancels
identically because the one-loop singular structure is proportional
to the Born tensor, whose T-odd projection vanishes. The T-odd
coefficient functions at this order therefore carry no explicit
renormalisation- or factorisation-scale dependence (the cross
sections depend on the scales only through $\as^2(\muR)$ and the
parton densities) --- a
property we use to validate the implementation --- and are given by six compact
closed forms of uniform weight one (rational functions and single
logarithms, no dilogarithms).

\subsection{Couplings and electroweak input}
\label{sec:formalism:couplings}

The electroweak couplings enter through channel-dependent bilinear
combinations of vector and axial couplings multiplying independent
PDF luminosities; the analytic implementation factorises every
coefficient function as a product of a colour average, a kinematic
structure function, an electroweak coupling factor, and a PDF
luminosity, with the full $\gamma^*/Z$ interference structure
in the neutral-current case and the CKM structure in the
charged-current case. Each helicity sector carries a
characteristic coupling bilinear (parity-even on both fermion
lines for $A_0$--$A_2$, parity-violating on both for $A_3$,
$A_4$, and mixed for the T-odd sector); the coupling conventions,
the sector structure, and the complete class-by-class luminosity
dictionary are collected in
Appendix~\ref{app:notation:couplings}. The implementation accepts
the standard electroweak input schemes; the specific scheme and
numerical inputs used for the results of this paper are given in
Section~\ref{sec:results:setup}.
The coupling and luminosity factors themselves were re-derived
independently from first principles rather than ported from the
existing code. This re-derivation exposed three genuine
flavour-structure defects in the inherited analytic $V{+}$jet
implementation of \dyturbo, which includes the axial-triangle
class in the $m_t\to\infty$ limit, with a
combined effect of $-3.8\times10^{-4}$ on the integrated
$V{+}$jet cross section: a missing isospin weight on the
luminosity combination of the axial-triangle class, a
flavour-summed combination where a flavour-weighted one is
required, and missing $Z$--$\gamma^*$ interference terms in two
coupling structures.

\section{Methodology}
\label{sec:methodology}

\subsection{The calculation chain}
\label{sec:methodology:chain}

All symbolic work was performed with open, scriptable software.
The derivation proceeds from Feynman diagrams to the compact
closed-form coefficient functions $C^\beta_{ab}(s,t,u,q^2,s_2)$ of
Eq.~\eqref{eq:master} through the following chain.

Diagrams for the one-loop virtual corrections and for the $2\to3$
real emission are generated with \form~5.0.1~\cite{Davies:2026cci}
and its multi-threaded variant
\textsc{TForm}~\cite{Tentyukov:2007mu}, using the integrated diagram
generator based on the \textsc{Grace} graph
generator~\cite{Kaneko:1994fd}; \form is also the main symbolic engine
for the Dirac and colour algebra, the helicity projections and the
$\epsilon$-expansions. The amplitudes are assembled in
conventional dimensional regularisation, $d = 4-2\epsilon$ (the
T-odd projections in strictly four dimensions,
Section~\ref{sec:formalism:projectors}), interfered with the Born
amplitudes, and projected on the helicity basis. The projection is
the only step at which the calculation distinguishes one angular
structure from another: the polarisation sum $-g_{\mu\nu}$ of the
unpolarised calculation is replaced by the covariant projector of
the desired structure
(Section~\ref{sec:formalism:projectors}), and every step that
follows is identical, so a single scalar run of the chain yields
every projection.

The one-loop integrals are reduced by integration-by-parts
identities to a basis of seven master integrals --- the massless
bubbles in $s$, $t$, $u$ and $q^2$ and the three one-mass boxes, the
triangles reducing to bubbles (Appendix~\ref{app:notation:virtual}) --- using \kira~3.1~\cite{Maierhofer:2017gsa,Klappert:2020nbg,Lange:2025fba} with
\firefly~2.0.3~\cite{Klappert:2019emp,Klappert:2020aqs} for the finite-field
rational reconstruction (\textsc{Fermat}~7.9~\cite{Lewis:Fermat} backend), accelerated
where needed with \ratracer~\cite{Magerya:2022hvj}. Their
$\epsilon$-expansions involve only a small fixed set of elementary
transcendental functions, referred to below as \emph{atoms}: the
logarithms $f_a$ (the index $a$ labels an atom here, not a parton
flavour) and the dilogarithm combinations
$f^{(1)}_{t,u}$, $f^{(2)}_{t,u}$ of
Appendix~\ref{app:coefficients}. The poles reproduce the
universal one-loop infrared structure quoted in
Section~\ref{sec:formalism:nlo}.

For the real emission the three-particle phase space is
factorised as
\begin{equation}
  \mathrm{d}\Phi_3
  = \frac{\mathrm{d}s_2}{2\pi}\;
    \mathrm{d}\Phi_2\bigl(s;\,q,K\bigr)\;
    \mathrm{d}\Phi_2^{*}\bigl(K;\,k_1,k_2\bigr) ,
  \qquad K^2 = s_2 ,
\label{eq:phi3}
\end{equation}
where $k_1$ and $k_2$ are the momenta of the two unobserved
final-state partons, $K = k_1 + k_2$ their total momentum, and
$\mathrm{d}\Phi_2^{*}$ the two-body phase space of the
recoiling parton pair in its rest frame, $\mathrm{d}\Phi_2^{*}
= \mathrm{d}\Omega^{*}/(32\pi^2)$ in four dimensions
($d$-dimensional wherever the angular integration produces
poles). The boson kinematics
$(q^2,\qT,y)$ fix $\mathrm{d}\Phi_2(s;q,K)$ completely, so the
entire nontrivial integration is over the recoil solid angle
$\Omega^{*}$, and it is performed analytically. On the two-body
cut every propagator is at most linear in the recoil decay
angles, the polar and azimuthal angles $(\theta_1,\theta_2)$ of the
recoil pair in its rest frame with
$\mathrm{d}\Omega^{*} = \mathrm{d}\cos\theta_1\,\mathrm{d}\theta_2$,
so after partial fractioning the angular integration closes on a
finite set of master integrals, whose closed forms, recursions and
constraint identities are collected in
Appendix~\ref{app:notation:angular}.

The result of this angular reduction is held in an intermediate
\emph{extended representation}: each
coefficient function $C^\beta_{ab}$ is a finite sum
\begin{equation}
  C^\beta_{ab}(s,t,u,q^2,s_2)
  = \sum_n c_n(s,t,u,q^2,s_2)\, I_n ,
\label{eq:extended}
\end{equation}
where $\beta$ labels the helicity projection and $ab$ the
partonic channel, the $I_n$ are the angular master integrals of
Appendix~\ref{app:notation:angular} --- the averages over the recoil
solid angle of products of inverse powers of the propagator
denominators of the two-body cut --- and the $c_n$ are rational
coefficients from the \form reduction. The compact closed forms
were then obtained from the extended
representation by inserting the closed forms of the basis angular
integrals and simplifying the result symbolically, with the
kinematic constraint $s_2 = s+t+u-q^2$ used to
eliminate one invariant (or, for the largest coefficient functions,
by finite-field reconstruction with \firefly/\ratracer followed by
partial fractioning), so the extended and the compact forms are one
mathematical object at two stages of simplification; comparisons
between them test this simplification step. The closed-form work at this
stage, the independent re-derivations, and the arbitrary-precision
numerical evaluations were carried out with
\sympy~1.14~\cite{Meurer:2017yhf} (with \texttt{mpmath}) and
\textsc{GiNaC}~1.8.10~\cite{Bauer:2000cp}.

The remaining $\epsilon$-dependence of the $s_2$ integration is
converted to distributions through the identity
\begin{equation}
  (s_2)^{-1-\epsilon}
  = (q^2)^{-\epsilon}\biggl\{
  -\frac{1}{\epsilon}\Bigl(\frac{A}{q^2}\Bigr)^{\!-\epsilon}\delta(s_2)
  + \left[\frac{1}{s_2}\right]_{A+}
  - \epsilon \left[\frac{\ln(s_2/q^2)}{s_2}\right]_{A+}
  + \order{\epsilon^2} \biggr\} ,
\label{eq:s2exp}
\end{equation}
valid on the slice $0\le s_2\le A$ over which the plus
distributions $[\,\cdot\,]_{A+}$ are defined
(Appendix~\ref{app:notation:phasespace}); the expansion of the
factor $(A/q^2)^{-\epsilon}$ generates the endpoint logarithm
$f_A = \ln(A/q^2)$, and $q^2$
is the reference scale of all logarithms. The identity
produces the closed $s_2$-distribution form of
Eq.~\eqref{eq:s2structure}; the poles then cancel as described in
Section~\ref{sec:formalism:nlo} --- the collinear poles at
$s_2\neq0$ against mass factorisation, point-wise in $s_2$, and the
endpoint $\delta(s_2)$ poles against the virtual corrections. The
resulting finite coefficient functions are listed in
Appendix~\ref{app:coefficients} and collected in the ancillary files.

\subsection{From closed-form coefficients to hadronic predictions}
\label{sec:methodology:workflow}

The implementation mirrors the $s_2$ structure of
Eq.~\eqref{eq:s2structure}: a one-dimensional endpoint sector
carrying the Born, virtual and $\delta(s_2)$ terms (and the entire
T-odd sector), integrated over the rapidity of the recoiling parton,
and a two-dimensional sector at $s_2\neq0$ carrying the real
emission with its plus distributions, integrated over a collinear
and a soft variable. In both, the parton momentum fractions are
fixed by the boson kinematics and by the integration variables, and
there the parton distributions are evaluated numerically through
\textsc{LHAPDF}~\cite{Buckley:2014ana} or
\textsc{NeoPDF}~\cite{Rabemananjara:2025unt}. The parametrisations, the
Jacobians and the integration limits are given in
Appendix~\ref{app:notation:convolution}; the integrations use
Gauss--Legendre quadrature on variables mapped so that the nodes
cluster at the endpoints. Among the many equivalent exact forms of
each coefficient function, the implementation uses the one that is
fastest to evaluate at the required accuracy: short polynomials in
the atoms and in a few rational denominators, computed once per
phase-space point and shared by all helicity cross sections, with
common subexpressions eliminated jointly across them.

The remaining outer integration over the boson kinematics
$(q^2,\qT,y)$ within experimental bins is performed with nested
sparse-grid (Smolyak) quadrature~\cite{Smolyak:1963,Gerstner:1998}
built on Gauss--Patterson rules~\cite{Patterson:1968};
the nested structure provides an embedded quadrature-error estimate
at no extra cost.

\subsection{Validation}
\label{sec:methodology:gates}

Every component had to pass the following verifications:
\begin{itemize}
\item \textbf{Structural identities}: Ward
  identities~\cite{Ward:1950xp,Takahashi:1957xn} per channel, i.e.\
  the vanishing of the hadronic tensor contracted with the boson
  momentum, $q_\mu H^{\mu\nu} = 0$, which current conservation
  requires;
  exact cancellation of all $1/\epsilon$ poles as symbolic
  identities; mass-dimension homogeneity scans of
  every emitted expression; $t\leftrightarrow u$ and beam-mirror
  parities; colour-scaling scans in $N_c$; the Lam--Tung relation
  $A_0=A_2$ at $\order{\as}$, channel by channel, on the emitted
  Born structure functions and their conversion to the \CS basis.
\item \textbf{Point-wise benchmarks}: matrix-element comparisons
  against \textsc{OpenLoops}~2.1.4~\cite{Buccioni:2019sur} and the
  \mcfm-derived code in \dyturbo at random
  phase-space points, in double, long-double and quadruple
  precision, typically at the $10^{-13}$--$10^{-15}$ level.
\item \textbf{Independent evaluation of the master integrals}: the
  masters and the multiple polylogarithms entering their
  $\epsilon$-expansions were re-evaluated numerically at high
  precision with \textsc{HepLib}~\cite{Feng:2021kha} using the
  auxiliary-mass-flow method of Ref.~\cite{Liu:2022chg}, with
  \textsc{pySecDec}~1.6.6~\cite{Borowka:2017idc}, and with
  \textsc{handyG}~0.1.5~\cite{Naterop:2019xaf}.
\item \textbf{Integrated closure}: the analytic $U{+}L$ result
  reproduces the established \dyturbo\ analytic $V{+}$jet module
  at $\order{10^{-14}}$ relative across all electroweak
  configurations, after the fixes described in
  Section~\ref{sec:formalism:couplings}; per-helicity integrated
  comparisons against the Monte Carlo campaign of
  Section~\ref{sec:results}.
\item \textbf{Asymptotics}: the $\qT\to0$ programme of
  Section~\ref{sec:lowqt}.
\end{itemize}
These verifications were applied continuously throughout the
development rather than once to the finished calculation, and
each fails in a characteristic pattern that localises the
defect. The structural identities act
before any number is produced: a wrong sign or a dropped colour
factor shows up as a residual pole or a broken
$t\leftrightarrow u$ parity in the symbolic expression itself. The
point-wise benchmarks localise a defect to a channel and a region
of phase space. The integrated comparison against the Monte Carlo
campaign is sensitive to what the point-wise tests cannot see,
namely a channel that is correct wherever it is evaluated but is
not evaluated everywhere it should be.

The comparison with Ref.~\cite{Mirkes:1992hu} confirmed its
$\order{\as^2}$ coefficient set, symbolically for every coefficient function
used, and isolated the
misprints collected in
Appendix~\ref{app:misprints}. The same comparison was made against
the four-quark classes of Ref.~\cite{Mirkes:1990diss}, which are not
part of the published literature and which no previous check could
use; they too are reproduced identically. The six T-odd coefficient functions
also agree, channel by channel and projection by projection, with
the structure functions of the independent recalculation of
Ref.~\cite{Lyubovitskij:2024civ} (which in turn reports agreement
with Refs.~\cite{Hagiwara:1984hi,Mirkes:1992hu}), up to constant
factors fixed by the projector and coupling conventions. The re-derivation of the luminosity
combinations from first principles exposed the flavour-structure
defects in the inherited analytic implementation that are listed at the end
of Section~\ref{sec:formalism:couplings}. Two further defects, in
the real-emission subtraction of the \mcfm-derived implementation
in \dyturbo, which descends from an early \mcfm version, were
exposed by the comparison with the Monte Carlo
campaign of Section~\ref{sec:results}. In the final-state
$g\to q\bar q$ splitting of the $V{+}$jet real radiation, the
Catani--Seymour dipole subtraction~\cite{Catani:1996vz} was built with the second beam
as the only spectator, while its integrated counterpart is
distributed over both beams; the mismatch is finite and free of
poles, it cancels in every $y$-even quantity, and it biases the
$y$-odd coefficient $A_1$ at $|y|>0$. Restoring the symmetric
spectator assignment yields a correction to $A_1$ of about
$7\times10^{-4}$ at $y>0$. In the
charged-current four-quark channels, the weights of the same-sign
dipole terms carry the CKM sum of the emitting quark twice, so that
the real emission is under-subtracted by the factor
$1-\sum_{q'}|V_{qq'}|^2$, of order $|V_{td}|^2\simeq8\times10^{-5}$
for down-type emitters. Both defects are absent from the current
\mcfm releases and are corrected in the numerical reference used
here. The
$\qT\to0$ closure of Section~\ref{sec:lowqt} exposed and localised
defects that every other verification in the list passed.

\subsection{Use of large language models}
\label{sec:methodology:llm}

The calculation was carried out with LLM agents of the Anthropic
Claude family (Claude Fable 5.1 and 5, Claude Opus 5.5, 5 and 4.8, and
Claude
Sonnet~5~\cite{Anthropic:2026fable5,Anthropic:2026fable51,Anthropic:2026opus48,Anthropic:2026opus5,Anthropic:2026opus55,Anthropic:2026sonnet5})
operating in the Claude Code environment~\cite{Anthropic:2026code} with shell
access to the software above, as the primary workforce for the
symbolic derivations, the code generation and implementation, the design
and execution of validation campaigns, and the project
documentation. Work was organised in bounded sessions (several
hundred over two months), each ending with a written hand-off note
recording claims, evidence and open items.
The human authors defined the physics goals, arbitrated standards
(including the acceptance thresholds),
reviewed the notes, and made all publication decisions.

\section{Results and validation against the numerical implementation}
\label{sec:results}

The primary validation of the analytic calculation is a full-phase-space
comparison against the numerical $V{+}$jet NLO implementation of
\dyturbo, whose matrix elements are inherited from
\mcfm~\cite{Campbell:1999ah,Campbell:2002tg} and integrated with
the adaptive Monte Carlo algorithm Vegas~\cite{Lepage:1977sw}. The two implementations share
only the electroweak couplings and the PDFs; the analytic result is an
independent derivation of the matrix elements, of their infrared
regularisation, and of the phase-space integration.

\subsection{Setup}
\label{sec:results:setup}

Four configurations are compared, chosen to match the binning of the
ATLAS angular-coefficient measurements:
$Z/\gamma^*$ production at 8~TeV in two rapidity regions,
$|y|<1.6$ ($z_{yc}$) and $1.6<|y|<3.6$ ($z_{yf}$), with
$80<m<100$~GeV and the 23 $\qT$ bins of
Ref.~\cite{ATLAS:2016rnf}; and $W^+$ and $W^-$ production at
13~TeV, rapidity-inclusive, with $50<m<150$~GeV and the 10
$\qT$ bins of Ref.~\cite{ATLAS:2025wai}. In all configurations the
$\qT$ integration starts at $\qTcut = 1$~GeV, the fixed-order
$V{+}$jet cross section being divergent as $\qT\to0$, so that the
first bin covers $\qTcut < \qT < 2.5$~GeV for $Z$ and
$\qTcut < \qT < 8$~GeV for $W^\pm$. Both sides use the
NNPDF4.0 NNLO PDF set~\cite{NNPDF:2021njg} with
$\as(m_Z) = 0.118$, and an electroweak
scheme in which all inputs are set explicitly: the couplings are
defined by the effective values
$\stw = 0.2293$ and $\alpha(m_Z) = 0.007779$ of
Ref.~\cite{Gauld:2017tww}, together with
$G_F = 1.1663787\times10^{-5}~\mathrm{GeV}^{-2}$,
$\mW = 80.385$~GeV and $m_Z = 91.1876$~GeV (the masses enter the
boson propagators with the fixed widths $\Gamma_Z = 2.4950$~GeV and
$\Gamma_W = 2.091$~GeV), and the CKM matrix elements
$|V_{ud}| = 0.97435$, $|V_{us}| = 0.22501$, $|V_{ub}| = 0.003732$,
$|V_{cd}| = 0.22487$, $|V_{cs}| = 0.97349$ and $|V_{cb}| = 0.04183$
for the charged current, with the massless bottom quark as an
initial-state parton and no top quark; the central scales are
$\muR = \muF = \sqrt{m^2+\qT^2}$. The axial-triangle contributions
are switched off on both sides for this comparison (the numerical
code does not include them); their effect is quantified in
Section~\ref{sec:triangles}.

The numerical reference was produced in a dedicated campaign:
the real-emission (VJ-real) and virtual (VJ-virt) terms were run as
separate tasks with $10^{10}$ and $10^{9}$ Vegas calls respectively,
with 1000 independent seeds per task (8000 jobs in total). The per-seed real-emission
distributions have heavy-tailed weights (excess kurtosis of $10$--$700$
per bin); the merged reference uses an outlier-trimmed combination for
the real term (per-bin removal of seeds farther than fifteen standard
deviations from the median, with the standard deviation estimated
from the central 68\% interval of the seed distribution) and a plain
average for the virtual term,
summing the two terms afterwards. The uncertainty assigned to each
numerical bin is the statistical error of the merged mean. For the T-odd coefficients
$A_5$--$A_7$ the numerator is taken from the virtual term alone: the
tree-level real emission and its dipole subtraction have zero mean for
the absorptive projections and contribute only Monte Carlo noise.

On the analytic side, the predictions are evaluated with
deterministic quadrature in the production setup timed in
Section~\ref{sec:costs} (with the $W$ rapidity
range split into three pieces), whose embedded error estimate is
below the Monte Carlo uncertainty in every bin; the entire
statistical uncertainty is assigned to the Monte Carlo side, and the
comparison metric is the per-bin pull
$(O^{\mathrm{ana}} - O^{\mathrm{num}})/\Delta O^{\mathrm{num}}$ for
each observable $O$, with $\Delta O^{\mathrm{num}}$ its numerical
uncertainty as defined above. The $\chi^2$ of an observable is formed
with the full bin-to-bin covariance of the numerical mean,
$\chi^2 = d^{\mathrm{T}} C^{-1} d$ with
$d = O^{\mathrm{ana}} - O^{\mathrm{num}}$: the seeds are independent,
so $C$ is estimated directly from their scatter, and the adaptive
integration correlates neighbouring $\qT$ bins appreciably.

\subsection{Comparison}
\label{sec:results:comparison}

\begin{figure}[htbp]
  \centering
  \includegraphics[width=0.98\textwidth]{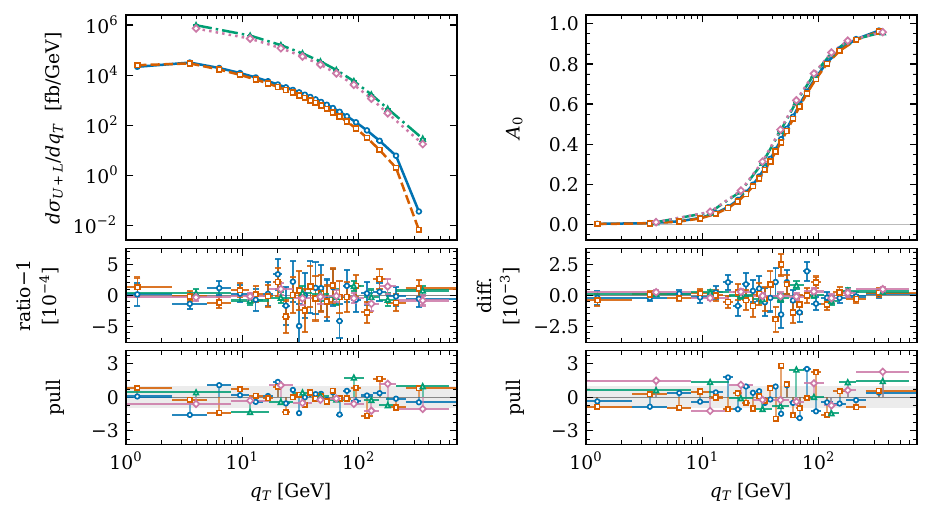}\\[1mm]
  \includegraphics[width=0.98\textwidth]{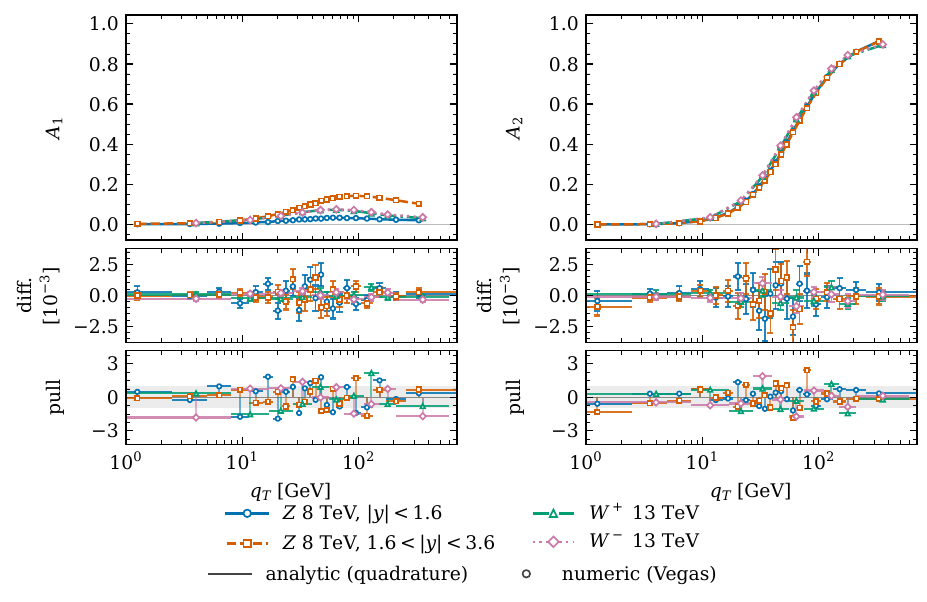}
  \caption{Analytic (lines) versus numerical (points with the total
    numerical uncertainty) results for $\sigUL$ and $A_0$ (top row) and for
    $A_1$, $A_2$ (bottom row), as functions of
    $\qT$, for $Z/\gamma^*$ at 8~TeV in the two rapidity
    bins $|y|<1.6$ and $1.6<|y|<3.6$ and for $W^{+}$, $W^{-}$ at
    13~TeV. For each coefficient the upper panel shows the
    observable, the middle panel the analytic$-$numerical difference
    (ratio for $\sigUL$), and the lower panel the pull, the analytic
    prediction being treated as exact.}
  \label{fig:anavsnum_even}
\end{figure}

\begin{figure}[htbp]
  \centering
  \includegraphics[width=0.98\textwidth]{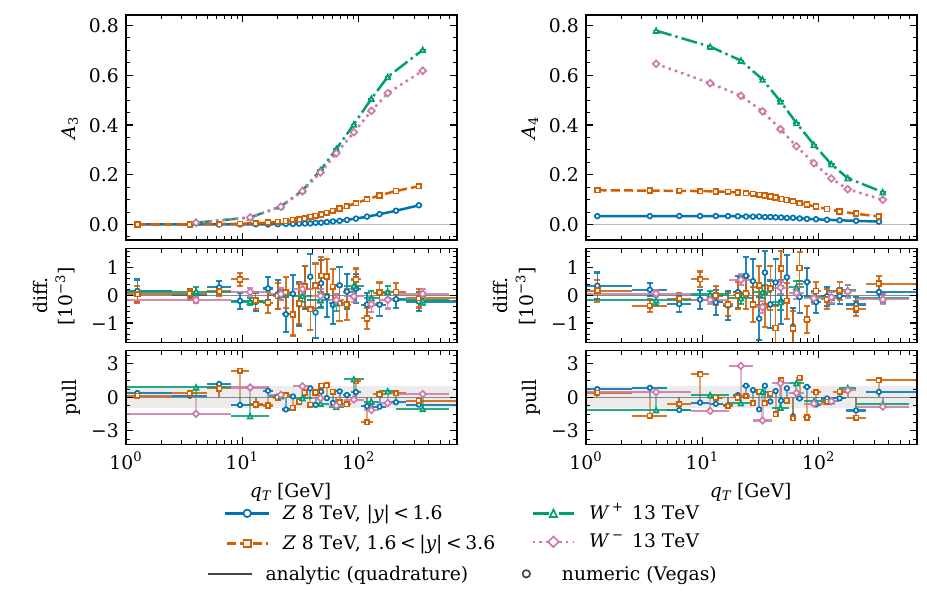}
  \caption{Same as Fig.~\ref{fig:anavsnum_even} for the
    parity-odd coefficients $A_3$, $A_4$.}
  \label{fig:anavsnum_odd}
\end{figure}

\begin{figure}[htbp]
  \centering
  \includegraphics[width=0.98\textwidth]{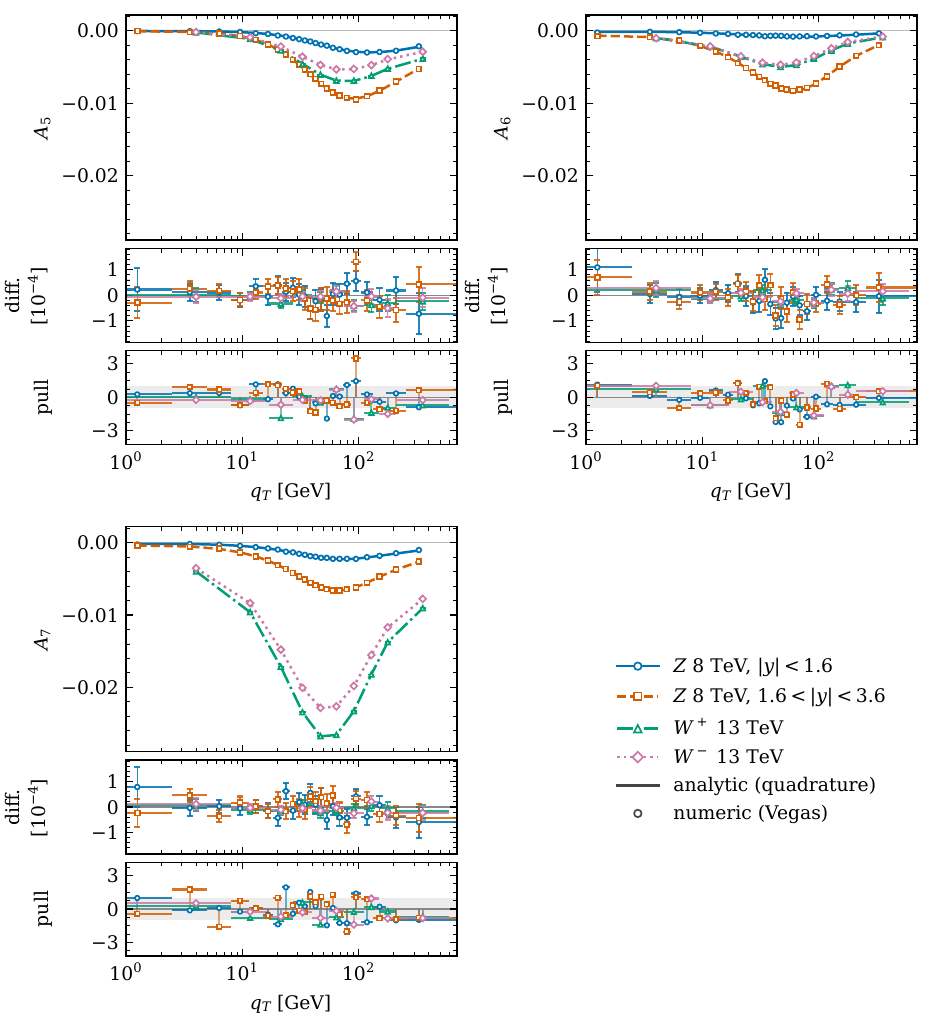}
  \caption{Same as Fig.~\ref{fig:anavsnum_even} for the T-odd
    coefficients $A_5$, $A_6$, $A_7$, shown in the internal sign
    convention of Section~\ref{sec:formalism:angular}, in which both
    sides of the comparison are computed.}
  \label{fig:anavsnum_todd}
\end{figure}

Table~\ref{tab:chi2} summarises the reduced $\chi^2$ of the
analytic-vs-numerical comparison for the unpolarised cross section
and the eight angular coefficients, for each configuration, together
with the global $p$-value of each configuration over its nine
observables. All 36 entries lie between
$\chi^2/\mathrm{ndf} = 0.45$ and $1.74$, and the global $p$-values are
$0.19$--$0.89$; given that the analytic side is treated as exact, this
establishes agreement between two NLO implementations that are
independent in their matrix elements, subtraction and phase-space
integration (they share the parton densities, the electroweak
couplings and the observable definition) at the level of the assigned
numerical uncertainty, which corresponds to a relative uncertainty on
$\sigUL$ of $0.8$--$4\times10^{-4}$ per bin ($2\times10^{-4}$ median)
in the $Z$ configurations and $0.6$--$1.3\times10^{-4}$ ($1\times10^{-4}$
median) for $W^\pm$, and to an absolute uncertainty on the $A_i$ between
$1\times10^{-5}$ and $2\times10^{-3}$ across bins and coefficients
($3$--$7\times10^{-4}$ median for $A_0$--$A_4$, $3$--$4\times10^{-5}$ for
$A_5$--$A_7$). The two largest entries, $W^-$ $A_1$ and
$W^+$ $A_0$ (both $1.74$), depend on the outlier trimming: with untrimmed
means of the same seeds they are $1.06$ and $0.98$.
Figures~\ref{fig:anavsnum_even}--\ref{fig:anavsnum_todd} show
the per-bin comparisons.
The comparison was repeated for variations of the renormalisation
and factorisation scales, and a similar agreement between the
numerical and the analytic calculations was found.

\begin{table}[htbp]
  \centering
  \caption{Reduced $\chi^2$ ($\chi^2/\mathrm{ndf}$) of the
    analytic-vs-numerical comparison for each
    helicity observable: the unpolarised cross section ($U{+}L$,
    absolute) and the angular coefficients $A_0$--$A_7$ (ratios), with
    the numerical uncertainty defined in the text (statistical uncertainty of
    the merged mean; $\chi^2$ with the full bin-to-bin covariance). The number of degrees of
    freedom is 23 for each $Z$ column and 10 for each $W$ column; the
    last two rows combine the nine observables of each column.}
  \label{tab:chi2}
  \begin{tabular}{lcccc}
    \toprule
     & $Z$, $|y|<1.6$ & $Z$, $1.6<|y|<3.6$ & $W^+$ & $W^-$ \\
    \midrule
    U$+$L   & 0.56 & 0.56 & 0.74 & 0.68 \\
    $A_0$   & 1.09 & 1.47 & 1.74 & 1.17 \\
    $A_1$   & 1.41 & 0.79 & 1.66 & 1.74 \\
    $A_2$   & 0.76 & 1.06 & 0.97 & 0.92 \\
    $A_3$   & 0.52 & 1.44 & 0.91 & 1.00 \\
    $A_4$   & 0.89 & 1.24 & 0.45 & 1.15 \\
    $A_5$   & 0.68 & 1.20 & 1.12 & 0.78 \\
    $A_6$   & 1.01 & 1.11 & 1.08 & 0.80 \\
    $A_7$   & 1.02 & 0.91 & 0.49 & 0.55 \\
    \midrule
    global $\chi^2/\mathrm{ndf}$ & 0.88 & 1.09 & 1.02 & 0.98 \\
    global $p$-value & 0.89 & 0.19 & 0.44 & 0.54 \\
    \bottomrule
  \end{tabular}
\end{table}

\section{Axial quark-triangle contributions}
\label{sec:triangles}

\subsection{Origin and structure}
\label{sec:triangles:structure}

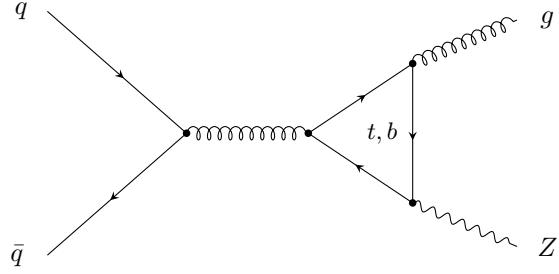
\begin{figure}[htbp]
  \centering
  \begin{tikzpicture}[scale=1.15, every node/.style={font=\small}]
    \coordinate (v)  at (0.8,0.0);
    \coordinate (t1) at (2.2,0.0);
    \coordinate (t2) at (3.4,0.8);
    \coordinate (t3) at (3.4,-0.8);
    \path[fermion] (-0.8,1.4) -- (v);   \node at (-1.1,1.4)   {$q$};
    \path[fermion] (v) -- (-0.8,-1.4);  \node at (-1.15,-1.4) {$\bar q$};
    \path[gluon]   (v) -- (t1);
    \path[fermion] (t1) -- (t2);
    \path[fermion] (t2) -- (t3);
    \path[fermion] (t3) -- (t1);
    \node[font=\footnotesize] at (3.05,0.0) {$t,b$};
    \path[gluon]   (t2) -- (4.6,1.3);   \node at (4.95,1.3)  {$g$};
    \path[vboson]  (t3) -- (4.6,-1.3);  \node at (4.95,-1.3) {$Z$};
    \node[vtx] at (v)  {};
    \node[vtx] at (t1) {}; \node[vtx] at (t2) {}; \node[vtx] at (t3) {};
  \end{tikzpicture}
  \caption{Representative diagram of the axial quark-triangle class in
    $q\bar q\to Zg$: the $Z$ boson attaches to a closed quark loop,
    connected to the open quark line by two gluons, one of which is
    the external on-shell gluon. Only the axial coupling of the $Z$
    to the loop survives, and only the top--bottom doublet contributes
    after the sum over flavours. The $qg\to Zq$ and $gq\to Zq$
    channels follow by crossing.}
  \label{fig:trianglediagram}
\end{figure}

At $\order{\as^2}$ a new gauge-invariant class of diagrams appears in
$Z{+}$jet production, in which the $Z$ boson attaches to a closed
quark loop connected to the open quark line by two gluons (one being
the external on-shell gluon). By Furry's theorem~\cite{Furry:1937zz} the vector coupling
of the closed loop vanishes, so only the axial coupling survives:
these are the anomaly-type triangle diagrams with one axial and
two vector couplings (AVV). Because a closed
quark loop cannot carry a single flavour-changing vertex, this class
contributes to neutral-current production only, with the $Z$ boson
attached to the loop (the photon, which has no axial coupling, enters
only through its interference with the $Z$ on the open quark line);
the $W^\pm$ helicity
cross sections receive no triangle contribution --- consistently, the
NLO $W$ calculation of Ref.~\cite{Mirkes:1992hu} contains no such
term.
A representative diagram is shown in Figure~\ref{fig:trianglediagram}.
These diagrams are part of the one-loop $Z{+}$jet amplitudes of
Ref.~\cite{Bern:1997sc}, given there in the large-$m_t$ expansion.
For the inclusive cross section their $\order{\as^2}$ contribution
was computed in
Refs.~\cite{Dicus:1985wx,Gonsalves:1991qn,Rijken:1995gi}, exact
top-mass one-loop amplitudes for $Z{+}$jet production were given in
Ref.~\cite{Campbell:2016tcu}, and the singlet axial-vector form
factor is known at three loops, in the massless
limit~\cite{Gehrmann:2021ahy} and with exact top-mass
dependence~\cite{Chen:2021rft}, entering the N$^3$LO Drell--Yan
predictions of Refs.~\cite{Neumann:2022lft,Campbell:2023lcy}.
What is new here is the complete set of helicity projections of the
massive triangle in closed form, and their implementation.
The $\qT$-differential unpolarised contribution in the
$m_t\to\infty$ limit is that of Ref.~\cite{Gonsalves:1989ar}.

Summing over the quark doublets, with all quarks but the top
massless, the degenerate light generations
cancel exactly and the triangle contribution reduces to the
top--bottom doublet,
\begin{equation}
  \sum_f a_f^{\mathrm{eff}}\, F(X; m_f)
  = \tfrac{1}{2}\left[ F(X; 0) - F(X; m_t) \right],
\end{equation}
where $a_f^{\mathrm{eff}} = -T_3^f$, with $T_3^f$ the weak isospin of
flavour $f$; the sign and the normalisation of the axial coupling
$a_f$ relative to $T_3^f$ are absorbed into $F$, so that the
relative sign between the two members of the doublet is the only
one that matters here;
where $F$ denotes the loop function of the relevant channel,
evaluated at the channel invariant $X = s, t, u$
for the $q\bar q$, $gq$, $qg$ channels respectively. Two distinct parts of the loop feed different helicity
structures: the dispersive part contributes to the T-even
sector ($\sigUL$ and $A_0$--$A_4$), while the absorptive
part feeds exclusively the T-odd coefficients $A_5$--$A_7$.

The absorptive projection is both ultraviolet and infrared finite,
so it can be computed in strictly four dimensions with an
anticommuting $\gamma_5$, with no scheme ambiguity and no
counterterm. The dispersive part is rendered unambiguous by imposing
the two vector Ward identities, placing the entire anomaly on the
axial index; the AVV tensor is then UV-finite with no residual
dependence on the regularisation scale. The resulting closed forms involve no dilogarithms: the
full mass dependence is carried by elementary logarithms and the
threshold variable $\beta_a = \sqrt{1-4m_f^2/a}$, where $m_f$ is
the mass of the quark in the loop and the invariant $a$ stands for
the channel invariant $X$ or the virtuality $q^2$.

\subsection{Exact top-mass dependence}
\label{sec:triangles:mt}

The $m_t\to\infty$ limit used in
Ref.~\cite{Gonsalves:1989ar} is a genuine power expansion in this
class:
with the vector Ward identities imposed, the top-quark part of the
tensor is $\order{s/m_t^2}$, with no logarithm and no constant
term surviving, so that only the $m_t$-independent massless
remnant is left in the limit --- unlike the singlet non-decoupling familiar from
inclusive Drell--Yan at higher orders, which requires the gluons to
be integrated over and would only reappear at NNLO in this
observable.

The exact top-mass dependence of the entire dispersive sector is
captured by a single scalar function,
\begin{equation}
  \sigma_i^{\mathrm{tri}}
  = \left[1 + \rho(X, q^2, m_t^2)\right]
    \sigma_i^{\mathrm{tri},\,m=0},
  \qquad
  \rho = - \frac{\Xi + q^2 (B_X - B_{q^2}) + m_t^2 (H_X - H_{q^2})}
               {\Xi - q^2 \ln|X/q^2|},
\label{eq:trirho}
\end{equation}
where $\sigma_i^{\mathrm{tri}}$ is the triangle contribution to
the helicity cross section $\sigma_i$, the index
$i \in \{U{+}L, 0, \ldots, 4\}$ runs over the dispersive (T-even)
projections, $\Xi = X - q^2$, and the auxiliary functions $B_a = \mathrm{Re}[\beta_a \ell_a]$,
$H_a = \mathrm{Re}[\ell_a^2]$,
$\ell_a = \ln\left[(\beta_a-1)/(\beta_a+1)\right]$ are evaluated
at $a = X$ and $a = q^2$, with the invariants understood as
$a + i0$, so that $\beta_a$ is real above threshold and imaginary
below it, and the real parts are taken after the continuation. This
closed form is continuous through both $2m_t$ thresholds and
reproduces the
independence of the mass correction from the helicity projection
to $2\times10^{-14}$. Its limits are
\begin{equation}
  \rho \;\xrightarrow{\;m_t\to\infty\;}\;
  \frac{\Xi^2}{12\, m_t^2 \left(\Xi - q^2\ln|X/q^2|\right)} ,
  \qquad
  \rho \;\xrightarrow{\;m_t\to0\;}\; -1 ,
\end{equation}
a clean $1/m_t^2$ approach with no logarithm on one side, and the
degenerate-doublet null on the other. The $\order{\as^2}$ real-emission partner of this
class --- the $q\bar q \to Z t\bar t$ axial interference --- is
included with exact $m_t$ in closed form as well; its physical
effect is below $10^{-5}$ of $\sigUL$ at LHC energies. The
calculation is carried out in the five-flavour scheme with a
massless bottom quark, for the boson distribution in $(q^2,\qT,y)$
integrated over the radiation, without any jet definition
(Section~\ref{sec:formalism}); the $t\bar t$ final state enters
only through this axial interference, the gauge-invariant partner
of the triangle class, and the remaining $Zt\bar t$ contributions,
which belong to heavy-quark-pair production rather than to the
massless $V{+}$jet calculation, are not included.

The calculation was validated point-wise against
\textsc{OpenLoops}~\cite{Buccioni:2019sur} (worst relative
deviation below $10^{-8}$ on the absorptive part, including the
below-threshold null and the $t\bar t$ threshold onset), against an
independent Cutkosky re-derivation in arbitrary-precision
arithmetic, and through vector Ward identities satisfied with no
counterterm at the $10^{-15}$ level.

\subsection{Phenomenological impact}
\label{sec:triangles:pheno}

\begin{figure}[htbp]
  \centering
  \includegraphics[width=0.98\textwidth]{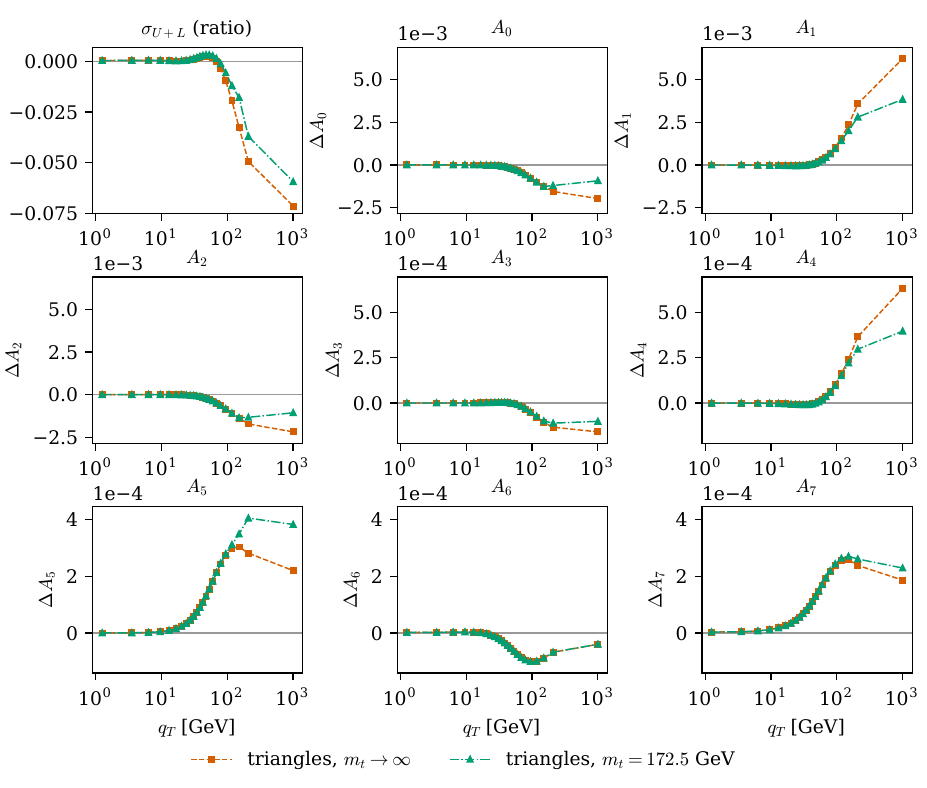}
  \caption{Effect of the axial triangle contributions on the
    $Z$-boson helicity observables at 8~TeV in the
    rapidity range $0<y<3.6$, for $m_t\to\infty$ and for finite
    $m_t=172.5$~GeV, relative to the calculation
    without triangles. Shown is the difference $\Delta A_i$ for each
    angular coefficient and the relative shift for $\sigUL$; the
    vertical scale is common within each sector ($A_0$--$A_2$,
    $A_3$--$A_4$, $A_5$--$A_7$), so the relative size of the effect
    can be compared across the coefficients of a sector. The
    $m_t\to\infty$ limit is shown for all observables.}
  \label{fig:triangles}
\end{figure}

Figure~\ref{fig:triangles} shows the effect of the triangle
contributions on the $Z$-boson angular coefficients at
8~TeV in the ATLAS binning, comparing triangles off,
triangles in the $m_t\to\infty$ limit, and triangles with finite
$m_t = 172.5$~GeV. All nine observables are shown in both mass
configurations.

The effect on $\sigUL$ is below the permille level everywhere
(reaching $-0.06\%$ in the highest $\qT$ bin). Among the T-even
coefficients the largest shift $\Delta A_i$ (the difference between
the prediction with and without the triangle contribution) is on
$A_1$, up to
$\Delta A_1 = 6\times10^{-3}$ in the $m_t\to\infty$ limit, reduced
to $4\times10^{-3}$ by the finite top mass; in the highest $\qT$ bin the
infinite-mass approximation overshoots the finite-mass result by up
to a factor $\sim2$. The relative effect on the T-odd coefficients
is much larger, since these are themselves loop-induced: at high
$\qT$ the triangle shifts $A_5$ and $A_7$ by $14\%$ and $17\%$
respectively. The $m_t\to\infty$ limit of
the absorptive triangle is the massless $qg$ remnant exactly: the
two massive cut contributions require $s > 4m_t^2$ and
$q^2 > 4m_t^2$ respectively, so once $4m_t^2$ exceeds
$\max(s, q^2)$ they switch off identically and the absorptive
part reaches its $m_t\to\infty$ limit at finite mass rather than
approaching it asymptotically (the dispersive part, in contrast,
approaches its limit as a power series in $s/m_t^2$). The resulting top-mass
sensitivity of the T-odd sector is small: relative to the
$m_t\to\infty$ limit, the finite top mass moves the $\qT$-integrated $\sigma_5$, $\sigma_6$ and
$\sigma_7$ by $4.5\times10^{-4}$, $1.5\times10^{-5}$ and
$2.5\times10^{-4}$, growing in the highest $\qT$ bin to $7\%$
and $4\%$ for $\sigma_5$ and $\sigma_7$ (where the $t\bar t$
thresholds open), while $\sigma_6$ peaks in the mid-$\qT$ range
at the $10^{-5}$ level.

\section{The low-\texorpdfstring{$\qT$}{qT} limit of the helicity cross sections}
\label{sec:lowqt}

\subsection{The validation at low \texorpdfstring{$\qT$}{qT}}
\label{sec:lowqt:def}

As $\qT\to0$ the fixed-order $V{+}$jet cross section diverges
logarithmically, and the divergent terms are predicted to all
logarithmic orders by transverse-momentum
factorisation. In a
$\qT$-subtraction or $\qT$-resummation matching scheme, the
fixed-order expansion of the resummed cross section --- the
\emph{asymptotic term}, in the nomenclature of
Refs.~\cite{Catani:2000vq,Bozzi:2005wk}, i.e.\ the complete set of
terms singular as $\qT\to0$ at this order, of the form
$\ln^k(q^2/\qT^2)/\qT$ for the cross sections differential in $\qT$ --- must
cancel the singular behaviour of the $V{+}$jet term point by point in
$(q^2, \qT, y)$. We call \emph{remainder} the difference between
the $V{+}$jet helicity cross section and the asymptotic term; it is
non-singular as $\qT\to0$, and it is the $V{+}$jet cross section
itself for the projections whose asymptotic term vanishes. We use
this cancellation as an internal test of the calculation, with both
terms evaluated at the same phase-space point.
The asymptotic term reproduces the complete leading-power behaviour
of the $V{+}$jet term, so that the remainder starts at the first
subleading power in $\qT^2/q^2$: the remainder multiplied by $\qT$
therefore follows a power law as $\qT\to0$, with logarithms as
slowly varying corrections. The test consists in reproducing that
power law over four decades in $\qT$.

The structure of the check differs across the helicity
projections. The asymptotic term is non-zero only for
$\sigUL$ and $\sigma_4$: the azimuthally symmetric $\qT\to0$ limit
populates only the azimuthally symmetric decay moments, and at
leading power the decay distribution is exactly $1+\cos^2\theta$
(times $\cos\theta$ for the parity-odd part), so that the
longitudinal projection vanishes as well (for $\sigma_4$, the
asymptotic term is the $U{+}L$ one with the symmetric Born
coupling weight replaced by the parity-odd one), while the
asymptotic terms of $\sigma_0$--$\sigma_3$ and of the T-odd projections
vanish identically. For $\sigUL$ and $\sigma_4$ the check therefore
probes a cancellation between two independently computed quantities
that grow like $\ln^k(\qT)/\qT$: at $\qT = 0.01$~GeV the remainder
is a factor $\sim3\times10^{7}$ smaller than either term. For the other seven projections the check instead verifies
that the analytic $V{+}$jet helicity cross sections are integrable
at $\qT \to 0$: the remainder multiplied by $\qT$ vanishes in the
limit, so that any singular behaviour of these projections is
weaker than $1/\qT$. This is the behaviour expected for the
subleading helicity structures, whose small-$\qT$ singularities are
integrable~\cite{Berger:2007si}, and it is where inverse powers of
$\qT$ introduced at intermediate stages of the derivation would
show up.

\subsection{Results}
\label{sec:lowqt:results}

\begin{figure}[htbp]
  \centering
  \includegraphics[width=\textwidth,height=0.44\textheight,keepaspectratio]{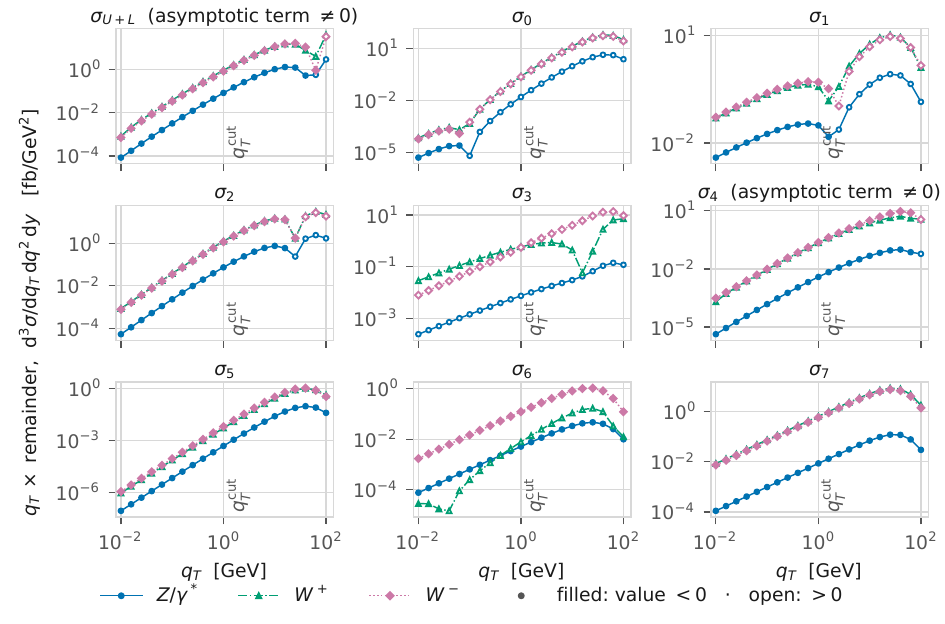}
  \caption{Low-$\qT$ behaviour of the $\order{\as^2}$ term of the
    remainder, multiplied by $\qT$, for the nine helicity projections
    $\mathrm{d}\sigma_i/(\mathrm{d}q^2\,\mathrm{d}\qT\,\mathrm{d}y)$, for $W^+$ and $W^-$ production at
    13~TeV ($m = \mW$) and $Z/\gamma^*$ production at 8~TeV ($m = m_Z$), at
    $y=1$. For $\sigUL$ and $\sigma_4$ this exhibits the
    cancellation between the $V{+}$jet cross section and the
    asymptotic term; for the remaining projections
    the asymptotic term vanishes and the curves demonstrate that
    the helicity cross sections are integrable at $\qT\to0$ ($\qT$
    times the remainder vanishes in the limit). The vertical line marks the cut $\qTcut = 1$~GeV of the
    predictions of Section~\ref{sec:results}.}
  \label{fig:lowqt}
\end{figure}

Figure~\ref{fig:lowqt} shows the $\order{\as^2}$ term of the remainder,
i.e.\ the $\order{\as^2}$ coefficient of the $V{+}$jet cross section
minus the $\order{\as^2}$ term of the asymptotic expansion, multiplied
by $\qT$, for all nine helicity projections, for $W^+$ and $W^-$ at
13~TeV and $Z/\gamma^*$ at 8~TeV, at $y=1$, over four
decades
$0.01 \le \qT \le 100$~GeV. All projections approach $\qT\to0$ as
clean power laws down to $\qT = \order{10^{-2}}$~GeV --- two
orders of magnitude below the cut $\qTcut = 1$~GeV of
Section~\ref{sec:results:setup}, marked by the vertical line.

The value of this limit as a development tool deserves emphasis:
probing the calculation far below its operating point exposed, and
localised, defects that were invisible to all other verifications
--- point-wise matrix-element comparisons, integrated cross-section
comparisons at percent-level Monte Carlo precision, and
machine-precision regression tests all passed while these defects
were present. All such defects were repaired at their source in the
symbolic derivation; what remains, and is described next, is the
numerical treatment needed to evaluate the correct expressions in
double precision over the four decades of Figure~\ref{fig:lowqt}.

\subsection{Numerical treatment of the limit}
\label{sec:lowqt:numerics}

The numerical difficulty of the $\qT\to0$ region has three sources.
The $V{+}$jet and asymptotic terms each grow like
$\ln^k(\qT)/\qT$ while their sum stays finite, so that every
relative error of the $V{+}$jet integral is amplified by the ratio
of the two, which reaches $3\times10^{7}$ at $\qT=0.01$~GeV for
$\sigUL$ and $\sigma_4$. The soft endpoint $s_2\to0$ and the
collinear endpoints of the real-emission integration are approached
at the same time. Finally, the Gram invariant
$\lambda^2 = (t+u)^2 - 4 s_2 q^2 = (t-u)^2 + 4\,s\,\qT^2$ vanishes on
the $t=u$ surface only as $\qT\to0$; the coefficient functions carry
inverse powers of $\lambda^2$ with cancelling residues, and their
direct evaluation near that surface loses most of the digits of
double-precision arithmetic, so that a single quadrature node can
carry most of an integral.

These difficulties are addressed as follows. The two quantities that
vanish in the limit, $s\,\qT^2 = tu - s_2 q^2$ and $\lambda^2$, are
computed directly from the phase-space variables, without
subtractions, and enter the coefficient functions as inputs wherever
they factor a numerator or a denominator, while the remaining
polynomials are homogenised in $(t,u,s_2,q^2)$ through the kinematic
constraint. The angular master integrals of the $qg$ channel, which
carry its transcendental content and are generated by a recursion
that divides by $\lambda^2$ at every step, are evaluated near the
$t=u$ surface from their convergent series in $\lambda^2$, whose
coefficients are exact rationals and which reaches double precision
with sixteen terms. The remaining coefficient functions that
degenerate on that surface are evaluated there either with
expansions about the surface or in quadruple precision, and the
parity-odd coefficient functions are evaluated in extended precision
as $s_2\to0$; kinematic conditions restrict these evaluations to a
small fraction of the nodes. The soft endpoint of the real-emission
integration is treated as a plus distribution, with the leading
$1/s_2$ and $\ln(s_2)/s_2$ behaviour taken from the exact collinear
pole, and below a switch point, chosen at each node by comparison
with the direct evaluation, the subtracted integrand is replaced by
its closed-form $s_2\to0$ expansion derived from the same
coefficient functions. Finally, the quadrature of the $V{+}$jet term
uses several times more nodes below $1$~GeV.

The residual double-precision error of this treatment was measured
by re-evaluating the $qg$ coefficient functions in quadruple
precision at every node of the same quadrature. Only the most
sensitive projection, $\sigma_2$, whose $qg$ contribution cancels
internally by four orders of magnitude at $\qT=0.01$~GeV, is
affected: its double-precision error is at the permille level at
$\qT=0.01$~GeV and below $10^{-5}$ above $\qT=0.04$~GeV, while the
other eight projections agree with the quadruple-precision
evaluation to better than $10^{-8}$ over the whole range.

\section{Computational cost}
\label{sec:costs}

\subsection{CPU cost: analytic versus numerical}
\label{sec:costs:cpu}

The computational cost of the analytic and of the numerical
calculations can be compared directly, since the two implementations
of Section~\ref{sec:results} produced the same predictions for
$Z$ and $W$ bosons, for all the helicity cross sections, once by
the Monte Carlo campaign and once by the analytic code.

The numerical campaign consumed $4.1\times10^{5}$ core-hours of
grid CPU --- a slot occupancy of $6.8\times10^{5}$ core-hours at
60\% CPU efficiency (Table~\ref{tab:cpu}) --- corresponding to $4.4\times10^{13}$ Vegas integrand
evaluations across 8000 jobs. The analytic predictions for the same
observables, with the quadrature of Section~\ref{sec:methodology} in
the setup of Table~\ref{tab:chi2} (with the $W$
rapidity range split into three pieces run as separate jobs),
required 0.86 core-hours --- a reduction by a factor
$4.7\times10^{5}$, with no residual statistical noise. The analytic
timings were measured on one performance core of an Intel Core
Ultra~7 155H processor, with the jobs pinned to that core and run
one at a time, while the grid jobs ran on the heterogeneous hardware
of the WLCG sites, whose per-core speed is not tracked; the
comparison is meant at the order-of-magnitude level. Expressed as
energy and carbon footprint in the model of
Ref.~\cite{Lannelongue:2021gag} (per-core power including the
memory share, at a carbon intensity of 250~g~CO$_2$e/kWh), the
numerical campaign corresponds to approximately
$3.2\times10^{3}$~kWh ($\sim0.8$~t~CO$_2$e), the
analytic one to $\sim0.007$~kWh ($\sim2$~g~CO$_2$e).

\begin{table}[htbp]
  \centering
  \caption{CPU cost of the four-configuration benchmark of
    Section~\ref{sec:results}.}
  \label{tab:cpu}
  \begin{tabular}{lrr}
    \toprule
     & numerical (grid) & analytic \\
    \midrule
    CPU time [core-h]        & $405\,486$ & $0.86$ \\
    integrand calls          & $4.4\times10^{13}$ & --- \\
    jobs                     & 8000 & 72 \\
    energy [kWh]             & $3175$ & $\sim0.007$ \\
    CO$_2$e [kg]             & $794$ & $\sim0.002$ \\
    \bottomrule
  \end{tabular}
\end{table}

The cost of the analytic code is dominated by the even ($A_0$--$A_2$,
58\%) and parity-odd ($A_3$, $A_4$, 36\%) NLO sectors; the T-odd
coefficients, being one-loop-absorptive only, are essentially free
(1\%), as is the unpolarised cross section (5\%). A single
helicity observable in a single configuration evaluates in seconds
to minutes on one core, making the analytic implementation suitable
for direct use inside profile-likelihood fits.

\subsection{LLM token budget}
\label{sec:costs:llm}

Because the use of LLM agents in a calculation of this scale is
novel, we document its cost with the same care as the CPU budget.
The natural unit of LLM usage is the token, the sub-word unit of
text in which language models read and generate text --- a few
characters, roughly three quarters of an English word, with similar
ratios for source code and formulas; billed usage counts the input
tokens read by the model, including repeated reads of cached
conversation context, and the tokens generated as output.
The accounting is derived from the complete session transcripts
of the project, which lasted approximately two months.
Table~\ref{tab:tokens} lists the resulting usage by model.

\begin{table}[htbp]
  \centering
  \caption{LLM usage attributed to this project, by model.
    ``Tokens'' counts all input tokens including cache reads; the
    cost is the list-price equivalent (the work was performed under
    a subscription).}
  \label{tab:tokens}
  \begin{tabular}{lrrrr}
    \toprule
    model & messages & tokens [$10^9$] & output [$10^6$] & list price [\$] \\
    \midrule
    Claude Fable 5.1 &  $3\,708$ &  1.46 &  3.7 & 1038 \\
    Claude Fable 5   & $11\,599$ &  3.20 &  7.5 & 4821 \\
    Claude Opus 5.5  &  $1\,607$ &  0.73 &  1.5 &  363 \\
    Claude Opus 5    & $38\,603$ & 10.34 & 14.9 & 8022 \\
    Claude Opus 4.8  &  $8\,389$ &  2.25 &  6.2 & 1605 \\
    Claude Sonnet 5  & $27\,734$ &  5.52 &  1.8 & 2174 \\
    \midrule
    total            & $91\,640$ & 23.50 & 35.6 & $18\,023$ \\
    \bottomrule
  \end{tabular}
\end{table}

Two observations are worth recording. First, 97\% of the input
tokens are prompt-cache reads --- the signature of long-context
agentic sessions --- which is what keeps the list-price equivalent
at $\approx\$1.8\times10^{4}$ rather than $\approx\$10^{5}$. Second, the
token cost is comparable to the CPU cost of the numerical
campaign: converted at
commercial-cloud prices --- between the $\approx\$0.014$ per
core-hour obtained by CMS on the Amazon spot market~\cite{CMS:2017keu}
and the $\approx\$0.045$ on-demand list price of current-generation
compute instances\footnote{Amazon EC2 on-demand pricing,
\url{https://aws.amazon.com/ec2/pricing/on-demand/}, September 2026;
the list-price basis consistent with the one used for the LLM cost.}
--- the $4.1\times10^{5}$ core-hours of the validation campaign alone
correspond to $\$(0.6$--$2)\times10^{4}$, comparable to the entire
LLM budget of the project.

\section{Summary and conclusions}
\label{sec:conclusions}

We have presented a complete analytic calculation of the NLO QCD
helicity cross sections for $W^\pm$ and $Z/\gamma^*$ production at
finite transverse momentum: the unpolarised cross section and the
eight helicity projections determining the angular coefficients
$A_0$--$A_7$ in the \CS frame, for both neutral- and charged-current
Drell--Yan, including the T-odd coefficients $A_5$--$A_7$ generated
by one-loop absorptive parts, and the axial quark-triangle
contributions with exact top-quark-mass dependence.

The results are implemented in \dyturbo and validated along three
independent lines: (i) against the established numerical
\mcfm-derived implementation, with $\chi^2/\mathrm{ndf}$ between
0.45 and 1.74 across all 36 configuration--observable entries, at
the few-times-$10^{-4}$ Monte Carlo precision of a dedicated
$4.1\times10^{5}$ core-hour campaign; (ii) through the analytic
$\qT\to0$ limit, where the singular behaviour cancels against the
asymptotic term of transverse-momentum resummation over four orders of
magnitude in $\qT$; and (iii) against the point-wise and structural
verifications of the derivation itself (Ward identities, exact symmetries,
extended-precision references, and the repaired analytic results of
the classic literature). Along the way, misprints
in the NLO angular-coefficient literature were identified
and corrected (Appendix~\ref{app:misprints}), including in the
four-quark classes that were never published.

The analytic implementation replaces $\order{10^{13}}$ Monte Carlo
integrand evaluations with deterministic quadrature over smooth
closed-form coefficient functions, reducing the CPU cost of a full
set of helicity predictions by more than five orders of magnitude
and eliminating statistical noise entirely. This makes NLO-accurate
angular coefficients --- including their scale and PDF variations
--- directly usable inside the profile-likelihood fits of precision
electroweak measurements such as $\mW$ and $\stw$. This work thus
closes a gap of more than thirty years between the original
analytic calculations of the angular coefficients and the release
of a public code for their evaluation in phenomenological
applications.

Finally, this work is, to our knowledge, among the first
perturbative QCD calculations of this scale carried out with LLM
agents as the primary workforce for derivation, implementation and
validation, under human direction and with machine-checkable
acceptance criteria for every result. The token budget of the entire
project ($\approx\$1.8\times10^{4}$ list-price equivalent) is comparable
to the CPU budget of its validation campaign; the methodology,
its verification strategy and its cost are documented in
Sections~\ref{sec:methodology} and~\ref{sec:costs} in the hope that
they are useful to other groups.

The natural continuation of this programme is the extension of the
same analytic chain to NNLO, where no closed-form literature exists
for the helicity cross sections; the present calculation was
structured throughout as a full-chain rehearsal for that step.

\section*{Acknowledgements}
We are indebted to Hannes Mildner and Florian Harz, who went to the library of the
Johannes Gutenberg University of Mainz and recovered
Ref.~\cite{Mirkes:1990diss} from microfilm.
LC is supported by Generalitat Valenciana plan GenT program
(CIDEGENT/2020/011). Part of the numerical calculations were carried out on
the Funes computing cluster, hosted at the computational infrastructure
of the Instituto de F\'isica Corpuscular (IFIC).
We acknowledge the computing resources that are provided by CERN.

\subsection*{Statement on the use of large language models}
In accordance with the publisher's and arXiv's policies on the use
of generative artificial-intelligence tools, we document that large
language models of the Anthropic Claude family were used in this
work as autonomous agents for symbolic derivation, code generation
and validation, under human direction and with machine-checkable
acceptance criteria for every result, as described in
Sections~\ref{sec:methodology:llm} and~\ref{sec:costs:llm}. No AI
system is an author of this work; the authors take full
responsibility for its entire content.

\clearpage
\appendix
\section{Notation and calculational framework}
\label{app:notation}

This appendix collects the notation and the framework formulas used
throughout the paper and in the ancillary files of
Appendix~\ref{app:coefficients}: kinematics, the helicity projectors
and their \CS conversion, the coupling and luminosity structure, the
phase-space factorisation, and the angular and loop master
integrals. The organising formula is
Eq.~\eqref{eq:master}, resolved into the
gauge-invariant classes $c$ of each channel,
$C^\beta_{ab}\,\mathcal{L}_{ab} = \sum_c C^\beta_{ab,c}\,\mathcal{L}_{ab,c}$:
\begin{equation}
  \frac{\mathrm{d}\sigma_i}{\mathrm{d}q^2\,\mathrm{d}\qT\,\mathrm{d}y}
  \;=\; \sum_{ab} \sum_{c} \int\!\mathrm{d}x_1\,\mathrm{d}x_2\;
  \mathcal{L}_{ab,c}(x_1,x_2;\muF)\;
  \sum_{\beta} M_{i\beta}\,
  C^{\,\beta}_{ab,c}(s,t,u,q^2,s_2;\as(\muR),\muF) .
\label{eq:lumassembly}
\end{equation}
The following subsections work out its pieces in turn: the
kinematics and the invariants on which the coefficient functions
depend (Appendix~\ref{app:notation:kinematics}), the projections
$\beta$ and the conversion $M_{i\beta}$ to the helicity cross
sections (Appendix~\ref{app:notation:projectors}), the luminosities
$\mathcal{L}_{ab,c}$ and the classes $c$
(Appendix~\ref{app:notation:couplings}), the $s_2$ distributions and
the phase space (Appendix~\ref{app:notation:phasespace}), the
variables of the $x_{1,2}$ integration
(Appendix~\ref{app:notation:convolution}), and the angular and loop
master integrals from which the coefficient functions are built
(Appendices~\ref{app:notation:angular} and
\ref{app:notation:virtual}); the coefficient functions
$C^\beta_{ab,c}$ themselves are provided as ancillary files
(Appendix~\ref{app:coefficients}).

\subsection{Kinematics}
\label{app:notation:kinematics}

Hadron- and parton-level Mandelstam variables for
$h_1(P_1)+h_2(P_2) \to V(q)+X$ with $p_i = x_i P_i$ are
\begin{equation}
\begin{aligned}
  S &= (P_1+P_2)^2 , & T &= (P_1-q)^2 , & U &= (P_2-q)^2 ,\\
  s &= (p_1+p_2)^2 = x_1 x_2 S , &
  t &= (p_1-q)^2 , & u &= (p_2-q)^2 ,\\
  t &= x_1 (T-q^2) + q^2 , & u &= x_2 (U-q^2) + q^2 , &&
\end{aligned}
\end{equation}
with $q^2 = m^2$ the lepton-pair virtuality, and
$\sqrt{S}$ the hadronic centre-of-mass energy. The invariant masses of the hadronic and partonic
recoil systems are
\begin{equation}
  S_2 = S + T + U - q^2 , \qquad
  s_2 = s + t + u - q^2 ,
\end{equation}
and the boson rapidity and transverse momentum are fixed by
\begin{equation}
  y = \frac{1}{2}\,
      \ln\!\left(\frac{q^2-U}{q^2-T}\right) , \qquad
  \qT^2 = \frac{(q^2-T)(q^2-U)}{S} - q^2 ,
\end{equation}
with the same relations at parton level,
$q^2 + \qT^2 = (q^2-t)(q^2-u)/s$. At fixed $(q^2,\qT,y)$ the
partonic invariants are functions of the momentum fractions alone;
in particular $s_2$ is linear in $x_2$ at fixed $x_1$, which is the
basis of the change of variables used for the real-emission
convolution, Eq.~\eqref{eq:x2tos2} below.

\subsection{Helicity structure and projectors}
\label{app:notation:projectors}

The nine helicity cross sections of
Eq.~\eqref{eq:angulardecomposition} are labelled
$\alpha \in \{U{+}L, L, T, I, P, A, 7, 8, 9\}$ in the internal
basis of Ref.~\cite{Mirkes:1992hu}, with harmonics
\begin{equation}
\begin{aligned}
  g_{U+L} &= 1+\cos^2\theta , &
  g_{L} &= 1-3\cos^2\theta , &
  g_{T} &= 2\sin^2\theta\cos2\phi ,\\
  g_{I} &= 2\sqrt{2}\,\sin2\theta\cos\phi , &
  g_{P} &= 2\cos\theta , &
  g_{A} &= 4\sqrt{2}\,\sin\theta\cos\phi ,\\
  g_{7} &= 2\sin^2\theta\sin2\phi , &
  g_{8} &= 2\sqrt{2}\,\sin2\theta\sin\phi , &
  g_{9} &= 4\sqrt{2}\,\sin\theta\sin\phi ,
\end{aligned}
\end{equation}
and the angular coefficients are the normalised ratios
\begin{equation}
\begin{aligned}
  A_0 &= \frac{2\,\sigma^{L}}{\sigUL} , &
  A_1 &= \frac{2\sqrt2\,\sigma^{I}}{\sigUL} , &
  A_2 &= \frac{4\,\sigma^{T}}{\sigUL} , &
  A_3 &= \frac{4\sqrt2\,\sigma^{A}}{\sigUL} ,\\
  A_4 &= \frac{2\,\sigma^{P}}{\sigUL} , &
  A_5 &= \frac{2\,\sigma^{7}}{\sigUL} , &
  A_6 &= \frac{2\sqrt2\,\sigma^{8}}{\sigUL} , &
  A_7 &= \frac{4\sqrt2\,\sigma^{9}}{\sigUL} .
\end{aligned}
\end{equation}

The parity-even projections are computed from the four raw
contractions of the hadronic tensor,
\begin{equation}
  I_{U+L} = -g_{\mu\nu}H^{\mu\nu}, \quad
  I_{L1}  = p_{1\mu}p_{1\nu}H^{\mu\nu}, \quad
  I_{L2}  = p_{2\mu}p_{2\nu}H^{\mu\nu}, \quad
  I_{L12} = \bigl(p_{1\mu}p_{2\nu}+p_{2\mu}p_{1\nu}\bigr)H^{\mu\nu},
\end{equation}
where terms proportional to $q^\mu$ have been dropped by current
conservation (enforced as exact Ward identities per channel).
With the normalisations
\begin{equation}
  \hat E_1^2 = \frac{(q^2-t)^2}{4q^2} , \qquad
  \hat E_2^2 = \frac{(q^2-u)^2}{4q^2} ,
\end{equation}
the projected structure functions in the $\beta$-basis are
\begin{equation}
  T^{\,U+L} = I_{U+L} , \qquad
  T^{\,L1} = \frac{I_{L1}}{\hat E_1^2} , \qquad
  T^{\,L2} = \frac{I_{L2}}{\hat E_2^2} , \qquad
  T^{\,L12} = \frac{I_{L12}}{\hat E_1 \hat E_2} ,
\end{equation}
and the conversion to the \CS helicity basis reads
\begin{equation}
  \begin{pmatrix} \sigma^{U+L}\\ \sigma^{L}\\ \sigma^{T}\\
                  \sigma^{I} \end{pmatrix}
  \propto
  \begin{pmatrix}
    1 & 0 & 0 & 0\\[1mm]
    0 & \dfrac{1}{4c_\gamma^2} & \dfrac{1}{4c_\gamma^2} & -\dfrac{1}{4c_\gamma^2}\\[2mm]
    \dfrac12 & -\dfrac{1+c_\gamma^2}{8s_\gamma^2 c_\gamma^2} & -\dfrac{1+c_\gamma^2}{8s_\gamma^2 c_\gamma^2} &
    \dfrac{1-3c_\gamma^2}{8s_\gamma^2 c_\gamma^2}\\[2mm]
    0 & \dfrac{1}{4\sqrt2\,\sqrt{s_\gamma^2 c_\gamma^2}} &
    -\dfrac{1}{4\sqrt2\,\sqrt{s_\gamma^2 c_\gamma^2}} & 0
  \end{pmatrix}
  \begin{pmatrix} T^{\,U+L}\\ T^{\,L1}\\ T^{\,L2}\\
                  T^{\,L12} \end{pmatrix},
\label{eq:csconversion}
\end{equation}
where
\begin{equation}
  c_\gamma^2 = \frac{sq^2}{(q^2-t)(q^2-u)} = \frac{q^2}{q^2+\qT^2}
      = \cos^2\gamma_{\mathrm{CS}} , \qquad s_\gamma^2 = 1-c_\gamma^2 ,
\end{equation}
is the (boost-invariant) cosine of the \CS beam angle. The matrix
of Eq.~\eqref{eq:csconversion} is that of Ref.~\cite{Mirkes:1992hu},
Eq.~(C.18), with the normalisation of its Eqs.~(C.13)--(C.14),
$\hat E_1^2 = (q^2-t)^2/(4q^2)$ and $\hat E_2^2 = (q^2-u)^2/(4q^2)$. In
the implementation the four coefficients on the basis
$(\hat g^{\mu\nu}, \hat p_1^\mu\hat p_1^\nu, \hat p_2^\mu\hat p_2^\nu,
\hat p_1^{\mu}\hat p_2^{\nu} + \hat p_2^{\mu}\hat p_1^{\nu})$, where
$\hat g^{\mu\nu} = g^{\mu\nu} - q^\mu q^\nu/q^2$ and
$\hat p_i^\mu = p_i^\mu - (p_i\!\cdot\!q/q^2)\,q^\mu$, so that
$\hat p_i^2 = -\hat E_i^2$, are obtained from the four contractions above
through the inverse of the $4\times4$ Gram matrix of the
basis (the parity-even part of $H^{\mu\nu}$ being symmetric,
$I_{L12} = 2\,p_{1\mu}p_{2\nu}H^{\mu\nu}$, which is how it is evaluated), generated symbolically in $d$ dimensions: its entries carry
the factor $1/(d-3)$ and the Gram determinant of $(p_1,p_2,q)$, and
the even-sector projection is therefore applied in $d = 4-2\epsilon$,
so that the $\epsilon$-dependence of the projector takes part in the
pole cancellation. In the runtime
implementation the \CS axes are built covariantly from the
unit spacelike projections of the beam momenta orthogonal to $q$,
\begin{equation}
  e_i^\mu = p_i^\mu - \frac{p_i\!\cdot\!q}{q^2}\,q^\mu , \qquad
  n_i^\mu = \frac{e_i^\mu}{\sqrt{-e_i^2}} , \qquad
  Z^\mu \propto n_1^\mu - n_2^\mu , \qquad
  X^\mu \propto n_1^\mu + n_2^\mu ,
\end{equation}
with $-e_i^2 = (p_i\!\cdot\!q)^2/q^2 = \hat E_i^2$, so that $Z$ and
$X$ are orthogonal ($n_1\!\cdot\!n_2$ cancels in $Z\!\cdot\!X$), both
normalised to $-1$ in the metric $(+,-,-,-)$; $Y$ is orthogonal to
the $p_1$--$p_2$--$q$ plane; the parity-odd
projections ($P$, $A$) are obtained from the dual vector
$d^\rho = \varepsilon^{\rho\sigma\mu\nu} \hat q_\sigma H^{A}_{\mu\nu}$,
with $\hat q = q/m$ and $H^{A}_{\mu\nu}$ the antisymmetric part of the
hadronic tensor, contracted with the $X$ axis for $\sigma_3$ and
with the $Z$ axis for $\sigma_4$; the T-odd projections involve the
axis normal to the event plane,
$Y^\mu \propto \varepsilon^{\mu\nu\rho\sigma}p_{1\nu}p_{2\rho}q_\sigma$,
through the $XY$, $ZY$ and $Y$ structures of $\sigma_5$, $\sigma_6$
and $\sigma_7$.

These projections are evaluated in the parton frame and mapped to
the \CS frame of Section~\ref{sec:formalism} inside the coefficient
evaluation, by an azimuthal rotation and, for the projections odd
under $z\to-z$ ($\sigma_1$, $\sigma_4$, $\sigma_5$, $\sigma_7$), the
sign that follows the boson longitudinal direction.

\subsection{Couplings and luminosities}
\label{app:notation:couplings}

The $Vq\bar q'$ vertex is written as
$-i\gamma^\mu(v_f - a_f\gamma_5)$, with $v_f = (g_L^f+g_R^f)/2$ and
$a_f = (g_L^f-g_R^f)/2$ in terms of the chiral couplings used by the
electroweak layer shared by all \dyturbo components,
\begin{align}
  Z:&\; g_L^f = g_Z\,(T_3^f - Q_f \stw),\quad g_R^f = -g_Z\, Q_f \stw ;
  \qquad
  \gamma:\; g_L^f = g_R^f = e\,Q_f ;
  \notag\\
  W:&\; g_L^{qq'} = \frac{g_W}{\sqrt2}\,V_{qq'},\quad g_R = 0 ,
\end{align}
the same forms holding for the leptons ($T_3 = -\tfrac12$, $Q=-1$,
unit mixing). In the $G_\mu$ scheme $g_W^2 = 4\sqrt2\,G_\mu m_W^2$ and
$g_Z^2 = g_W^2/\cos^2\theta_W = 4\sqrt2\,G_\mu m_Z^2$; in the scheme used for the results of
this paper (Section~\ref{sec:results}), which takes $\stw$ and
$\alpha$ as inputs, $g_W^2 = 4\pi\alpha/\stw$ and
$g_Z = g_W/\cos\theta_W$ (the electroweak layer of the code stores
one half of these chiral couplings and compensates the factor in
the vertex; the bilinears of Eq.~\eqref{eq:couplingstructure} are
unaffected). Below we write $L^V_{ff'} \equiv g_L^{ff'}$ and
$R^V_{ff'} \equiv g_R^{ff'}$ for the chiral couplings of the boson
$V$ to the flavour pair $(f,f')$, which is flavour diagonal for the
neutral current, and drop the label $V$ when no ambiguity arises.
Each helicity sector picks out a
characteristic bilinear structure in the lepton and quark
couplings:
\begin{equation}
\begin{aligned}
  \sigma^{U+L,L,T,I} &\;\propto\; (v_\ell^2+a_\ell^2)(v_q^2+a_q^2) ,\\
  \sigma^{P,A}       &\;\propto\; v_\ell a_\ell\, v_q a_q ,\\
  \sigma^{7,8}       &\;\propto\; (v_\ell^2+a_\ell^2)\, v_q a_q ,\\
  \sigma^{9}         &\;\propto\; v_\ell a_\ell\,(v_q^2+a_q^2) ,
\end{aligned}
\label{eq:couplingstructure}
\end{equation}
i.e.\ the T-even coefficients $A_0$--$A_2$ are parity-even on both
lines, the parity-odd $A_3$, $A_4$ require parity violation on
both, and the T-odd sector splits: $A_5$, $A_6$ carry the
parity-violating quark coupling with the parity-even lepton trace,
while $A_7$ is the reverse. For the neutral current every quark
bilinear is accompanied by the $\gamma\gamma$, $ZZ$ and
interference propagator factors; for the charged current the CKM
matrix enters the flavour sums.

Each luminosity of Eq.~\eqref{eq:lumassembly} is a sum over the flavours $f_1$, $f_2$ of the incoming
partons of the parton densities weighted by the electroweak factor
of the class,
\begin{equation}
  \mathcal{L}_{ab,c}(x_1,x_2;\muF)
  = \sum_{f_1 f_2} f_{f_1}(x_1,\muF)\, f_{f_2}(x_2,\muF)
    \sum_{V,V'} P_{VV'}(q^2)\;\ell^{\,\alpha}_{VV'}\;
    K^{c,\alpha}_{f_1 f_2;VV'} ,
\label{eq:lumidef}
\end{equation}
where $P_{VV'}$ are the propagator factors of the boson pair
($\gamma\gamma$, $ZZ$ and $\gamma Z$ for the neutral current, $WW$
for the charged current), $\ell^{\alpha}_{VV'}$ is the lepton
bilinear of Eq.~\eqref{eq:couplingstructure} for the helicity sector
$\alpha$ with the couplings of $V$ and $V'$ (e.g.\ $v_\ell^V
v_\ell^{V'} + a_\ell^V a_\ell^{V'}$ for the parity-even sectors), and
$K^{c,\alpha}$ is the quark coupling factor of the class, the
corresponding quark bilinear built from $L_{ff'}$ and $R_{ff'}$. For
the two-parton channels there is one class per projection, with
$K_q = L_q L'_q + R_q R'_q$ for the projections whose quark bilinear
in Eq.~\eqref{eq:couplingstructure} is $v_q^2+a_q^2$ and
$L_q L'_q - R_q R'_q$ for those with $v_q a_q$, $\sigma_3$--$\sigma_6$
(the primed couplings those of $V'$), and the luminosities read
\begin{equation}
  \mathcal{L}_{q\bar q} = \sum_q K_q\,
    \bigl[ f_q(x_1) f_{\bar q}(x_2) \pm f_{\bar q}(x_1) f_q(x_2) \bigr] ,
  \qquad
  \mathcal{L}_{qg} = \sum_q K_q\,
    \bigl[ f_q(x_1) \pm f_{\bar q}(x_1) \bigr] f_g(x_2) ,
\end{equation}
with the minus sign for $\sigma_3$--$\sigma_6$, the projections odd
under charge conjugation of the quark line (not under the interchange
of the beams), $\mathcal{L}_{gq}$ following from $\mathcal{L}_{qg}$
with the beams exchanged, the lepton and propagator factors of
Eq.~\eqref{eq:lumidef} understood; for the charged current $K_q$
carries $|V_{qq'}|^2$ and the $qg$ channel $\sum_{q'}|V_{qq'}|^2$. For
the four-quark channels the classes $c$ of Eq.~\eqref{eq:lumassembly}
are the squares and interferences of the four attachments of the
boson, each labelled by the quark line it sits on: the annihilating
line ($i$) or the produced pair ($p$) in the topology with the
$s$-channel gluon, and the beam-1 ($1$) or beam-2 ($2$) quark line
in the topology with the $t$-channel gluon. The coefficient
functions $C^\beta_{ab,c}$ are the $C^{\beta,\cdot}_{q\bar q,ij}$ of
$q\bar q \to V q'\bar q'$ and the $C^{\beta,\cdot}_{qq',ij}$ of
$qq' \to V qq'$ given in the ancillary files (Appendix~\ref{app:coefficients}), and
the luminosities $\mathcal{L}_{ab,c}$ are Eq.~\eqref{eq:lumidef}
with $f_{f_1}$, $f_{f_2}$ the densities of the two incoming partons.
The coupling factor of a class follows the flavour of the line the
boson attaches to: with
$B_f \equiv L_{ff}L'_{ff} + R_{ff}R'_{ff}$ the parity-even quark
bilinear of flavour $f$ (the primed couplings those of $V'$, as in
$K_q$ above), $f_1$ and $f_2$ the flavours of the two incoming
partons and $\delta_{f_1f_2}$ restricting them to a flavour-diagonal
$q\bar q$ pair, the neutral-current classes carry
\begin{equation}
\begin{aligned}
  ii,\,i1,\,i2 &: \;\delta_{f_1f_2}\,B_{f_1} , &
  ip &: \;\delta_{f_1f_2}\,(L_{f_1f_1}-R_{f_1f_1})
             \textstyle\sum_f (L_{ff}-R_{ff}) ,\\
  p1,\,p2 &: \;\delta_{f_1f_2}\,B_{f_1} , &
  pp &: \;\delta_{f_1f_2}\,\textstyle\sum_f B_f ,\\
  11 &: \;B_{f_1} , \qquad 22 : \;B_{f_2} , &
  12 &: \;L_{f_1f_1}L_{f_2f_2}+R_{f_1f_1}R_{f_2f_2}\;(LL) ,
\end{aligned}
\label{eq:lumiclasses}
\end{equation}
and $L_{f_1f_1}R_{f_2f_2}+R_{f_1f_1}L_{f_2f_2}$ for the $LR$ half of
the class $12$; the classes with the boson on a single line
($ii$, $i1$, $i2$, $p1$, $p2$, $11$, $22$) carry the coupling of
that line's flavour, the class $pp$ the sum over the produced
flavour, the class $ip$ the closed-flavour-loop sum, and the class
$12$ one chiral current per line. The classes without a
$\delta_{f_1f_2}$ are those of the scattering topology, where the
two incoming flavours are independent. These
factors are those of Eqs.~(2.18) and (2.21) of
Ref.~\cite{Gonsalves:1989ar};
two of them differ from the flavour bookkeeping inherited from the
numerical implementation and were corrected in the course of this
work: the class $ip$ carries the weight
$(L_{f_1f_1}-R_{f_1f_1})\sum_f(L_{ff}-R_{ff})$,
i.e.\ the weak-isospin sign $2T_3$ of the annihilating flavour times the sum over
the closed loop (which equals $-1$ for five flavours), rather than a
flavour-blind factor, and the class $p1$ carries the coupling of the
flavour of its own line rather than the flavour-summed one. The
charged-current four-quark classes carry $|V_{qq'}|^2$ for the
CKM-allowed flavour pairs. The axial-triangle luminosities of
Section~\ref{sec:triangles} carry $(L_{f_1f_1}-R_{f_1f_1})\sum_f
(L_{ff}-R_{ff})$ in the $qg$ and $gq$ channels and the same
$T_3$-weighted combination in the $q\bar q$ channel. Every factor
is built from the same electroweak primitives as the resummed part
of \dyturbo, which makes the fixed-order and resummed pieces
consistent in every electroweak scheme.

\subsection{Phase space and distributions}
\label{app:notation:phasespace}

The three-particle phase space of the real emission
$p_1+p_2 \to q + k_1 + k_2$ is factorised as in
Eq.~\eqref{eq:phi3}. For the PDF convolution, at fixed $x_1$ the variable $s_2$
replaces $x_2$~\cite{Gonsalves:1989ar},
\begin{equation}
  \int_0^1\!\mathrm{d}x_1\,\mathrm{d}x_2\;\theta(s_2)
  = \int_{x_1^{\mathrm{min}}}^{1}
    \frac{\mathrm{d}x_1}{x_1 S + U - q^2}
    \int_0^{s_2^{\mathrm{max}}}\!\mathrm{d}s_2 ,
  \qquad
  \begin{aligned}
  s_2^{\mathrm{max}} &= U + x_1 (S_2 - U) ,\\
  x_1^{\mathrm{min}} &= \frac{-U}{S_2-U} ,
  \end{aligned}
\label{eq:x2tos2}
\end{equation}
and the soft/collinear singularities at $s_2\to0$ are organised as
the distributions of Eq.~\eqref{eq:s2structure} with the
prescription
\begin{align}
  \int_0^{A}\!\mathrm{d}s_2\, f(s_2)
  \left[\frac{1}{s_2}\right]_{A+}
  &= \int_0^{A}\!\mathrm{d}s_2\,
    \frac{f(s_2)-f(0)}{s_2} ,\\
  \int_0^{A}\!\mathrm{d}s_2\, f(s_2)
  \left[\frac{\ln(s_2/q^2)}{s_2}\right]_{A+}
  &= \int_0^{A}\!\mathrm{d}s_2\,
    \frac{[f(s_2)-f(0)]\ln(s_2/q^2)}{s_2} .
\end{align}
The expansion producing them is the distributional identity of
Eq.~\eqref{eq:s2exp}, after which the $s_2>0$ collinear poles cancel point-wise in $s_2$
against the mass-factorisation counterterm
\begin{equation}
  \mathrm{d}\sigma^{\mathrm{MF}}
  = -\frac{1}{\epsilon}\,\frac{\as}{2\pi}\,
    \Bigl(\frac{4\pi\mu^2}{\muF^2}\Bigr)^{\!\epsilon}
    \frac{1}{\Gamma(1-\epsilon)}\;
    P^{(0)}_{ab} \otimes \mathrm{d}\sigma^{\mathrm{LO}} ,
\end{equation}
in the $\overline{\mathrm{MS}}$ scheme, with the four-dimensional
leading-order kernels $P^{(0)}_{ab}$ and the conventional factor
$(4\pi)^\epsilon/\Gamma(1-\epsilon)$ absorbed into the pole, and the endpoint
$\delta(s_2)$ poles cancel against the virtual corrections, whose
singular part is the universal one-loop structure of
Ref.~\cite{Catani:1998bh}, which for the
$q\bar q\to Vg$ channel reads
\begin{equation}
  \frac{2\,\mathrm{Re}\,\mathcal{M}_0^*\mathcal{M}_1}{|\mathcal{M}_0|^2}
  = \frac{\as}{2\pi}\Bigl[
    -\frac{2C_F+N_c}{\epsilon^2}
    -\frac{3C_F+\gamma_g}{\epsilon}
    -\frac{(N_c-2C_F)\,L_s - N_c(L_t+L_u)}{\epsilon}
    \Bigr] + \order{\epsilon^0},
\end{equation}
with $\gamma_g = \frac{11}{6}N_c - \frac{2}{3}T_R n_f$,
$L_x = \ln(|x|/\mu^2)$ and $\mu$ the dimensional-regularisation scale
(the imaginary part of the logarithm of the timelike invariant
does not contribute to the real part of the interference at pole
level); the $qg$ channel follows by crossing.
The poles of the reduced virtual amplitude were verified to reproduce
this structure exactly as rational functions of $t$, $q^2$ and $N_c$,
and the finite remainder was adjudicated against \mcfm.

\subsection{Numerical evaluation of the PDF convolution}
\label{app:notation:convolution}

The endpoint ($\delta(s_2)$) and the $s_2>0$ sectors of
Eq.~\eqref{eq:master} are integrated as follows.
At the endpoint, $\delta(q^2-s-t-u)$ removes one
integration and the remaining one is taken over the rapidity $y_3$
of the recoiling parton. With $m_T^2 = m^2+\qT^2$, the momentum
fractions of Eq.~\eqref{eq:factorization} are fixed to
\begin{equation}
  x_1 = \frac{m_T e^{y} + \qT e^{y_3}}{\sqrt{S}} ,
  \qquad
  x_2 = \frac{m_T e^{-y} + \qT e^{-y_3}}{\sqrt{S}} ,
\label{eq:xdelta}
\end{equation}
so that
\begin{equation}
  \sigma^{\delta}
  = \int_{y_3^{\min}}^{y_3^{\max}}\!\!\mathrm{d}y_3\;
    J(y_3) \sum_{ab} f_a(x_1,\muF)\, f_b(x_2,\muF)\;
    C^{\delta}_{ab} ,
\label{eq:sigmadelta}
\end{equation}
with $J(y_3) = 1/S$ the Jacobian of this parametrisation: at
fixed $x_1$, $\partial s_2/\partial x_2 = \sqrt{S}\,\qT e^{y_3}$ and
$\mathrm{d}x_1/\mathrm{d}y_3 = \qT e^{y_3}/\sqrt{S}$, so that
$\mathrm{d}x_1\,\mathrm{d}x_2\,\delta(s_2) = \mathrm{d}y_3/S$;
$C^{\delta}_{ab}$ the endpoint coefficient function of channel
$ab$ --- the coefficient $C^{\beta,\delta}_{ab}$ of $\delta(s_2)$ in Eq.~\eqref{eq:s2structure},
carrying the Born, virtual and endpoint contributions --- and the
limits $y_3^{\max} = \ln[(\sqrt{S}-m_T e^{y})/\qT]$ and
$y_3^{\min} = -\ln[(\sqrt{S}-m_T e^{-y})/\qT]$ set by $x_1\le1$ and
$x_2\le1$. This integral is performed with Gauss--Legendre
quadrature.

Away from the endpoint the recoiling partons carry an invariant
mass $s_2>0$, which frees a second integration. Writing
$x_1^0 = (m_T/\sqrt{S})\,e^{y}$ and
$x_2^0 = (m_T/\sqrt{S})\,e^{-y}$ for the endpoint fractions, the
two variables are a collinear variable $z_1$ and a soft variable
$z_2$ carrying the $s_2$ dependence:
\begin{equation}
  x_1 = \frac{x_1^0\,(1+\xi)}{1-z_2} ,
  \quad
  \xi = \frac{\qT^2}{m_T^2}\,\frac{1-z_1}{z_1} ,
  \qquad
  x_2 = \frac{x_2^0}{1-z_1} ,
  \qquad
  s_2 = z_1 z_2\,s
      = \frac{z_1 z_2\, m_T^2\,(1+\xi)}{(1-z_1)(1-z_2)} ,
\label{eq:xreal}
\end{equation}
with $z_1 \in [z_1^{\min},\,1-x_2^0]$,
$z_1^{\min} = x_1^0\qT^2/[x_1^0\qT^2+(1-x_1^0)m_T^2]$, set by
$x_1\le1$ and $x_2\le1$ at $z_2=0$, and $z_2 \in
[0,\,1-x_1^0(1+\xi)]$, set by $x_1\le1$; $z_2\to0$ is the soft
endpoint $s_2\to0$ at fixed $z_1$, and $z_1\to1$ the collinear
limit. The sector then reads
\begin{equation}
  \sigma^{s_2>0}
  = \int\!\mathrm{d}z_1\,\mathrm{d}z_2\;
    J(z_1,z_2) \sum_{ab} f_a(x_1)\, f_b(x_2)\;
    \bigl[ C_{ab}(s_2) \bigr]_{+} ,
\label{eq:sigmareal}
\end{equation}
where
\begin{equation}
  J(z_1,z_2) = \biggl|\frac{\partial(x_1,x_2)}{\partial(z_1,z_2)}\biggr|
  = \frac{x_1^0 x_2^0\,(1+\xi)}{[(1-z_1)(1-z_2)]^2}
  = \frac{m_T^2}{S}\,\frac{1+\xi}{[(1-z_1)(1-z_2)]^2}
\end{equation}
is the Jacobian of the $(z_1,z_2)$ parametrisation
($\partial x_2/\partial z_2 = 0$, so the determinant is
$-\,\partial x_1/\partial z_2\;\partial x_2/\partial z_1$) and $[\;]_{+}$
denotes the plus-distribution structure of Eq.~\eqref{eq:s2structure}
acting, at fixed $z_1$, on the $z_2$ dependence of the product of
coefficient function and parton densities, with $s_2 = z_1 z_2s$
and the upper limit $A$ of the $s_2$ slice at $z_2 = A/(z_1s)$. Both integrations use
Gauss--Legendre quadrature on variables mapped so that the nodes
cluster at the endpoints: a logit map for $z_1$, whose integrand is
peaked at the collinear boundaries, and an exponential map for
$z_2$, whose integrand behaves as $(c_1\ln z_2 + c_0)/z_2$, with
$c_0$, $c_1$ constants, as $z_2\to0$, and as $c_1\ln z_2 + c_0$
after the plus subtraction. The implementation evaluates the
momentum-weighted densities $x f_a(x,\muF)$ returned by the PDF
libraries, and accordingly carries the Jacobians $1/s$ and
$m_T^2/s$, which differ from those above by $1/(x_1x_2)$.

\subsection{Angular master integrals}
\label{app:notation:angular}

On the two-body cut every propagator denominator is at most linear
in the recoil decay angles,
\begin{equation}
  D_k = \mathcal{A}_k + \mathcal{B}_k\cos\theta_1
        + \mathcal{C}_k\sin\theta_1\cos\theta_2 ,
\label{eq:angulardenom}
\end{equation}
with process-dependent constants
$(\mathcal{A}_k,\mathcal{B}_k,\mathcal{C}_k)$ for each propagator
family $k$, satisfying
$\mathcal{B}_k^2+\mathcal{C}_k^2 = \lambda^2/4$ and
$\mathcal{A}_k^2-\mathcal{B}_k^2-\mathcal{C}_k^2 = sq^2$, where
$\lambda^2 = (t+u)^2 - 4 s_2 q^2$ is the Gram combination that
controls the $t=u$ surface discussed in
Section~\ref{sec:lowqt}. The angular master integrals $I_n$ of
Eq.~\eqref{eq:extended} are the averages
\begin{equation}
  I_n = \frac{1}{4\pi}\int\!\mathrm{d}\Omega^{*}\;
        \frac{(-\cos\theta_1)^{j}}{\prod_{k} D_k^{\,i_k}} ,
\label{eq:angularbasis}
\end{equation}
with one factor per propagator family of the two-body cut, and they
are expressed in the single-denominator classes
\begin{equation}
  I^{(i,j)}_k = \frac{1}{4\pi}\int\!\mathrm{d}\Omega^{*}\;
        \frac{(-\cos\theta_1)^{j}}{D_k^{\,i}} ,
\label{eq:angularclass}
\end{equation}
in which the coefficient functions of the ancillary files
(Appendix~\ref{app:coefficients}) are written. The two base integrals
are (with $f_\lambda$ as listed in
Appendix~\ref{app:coefficients})
\begin{equation}
  I^{(1,0)}_k = \frac{1}{2\mathcal{R}_k}\,
      \ln\frac{\mathcal{A}_k+\mathcal{R}_k}{\mathcal{A}_k-\mathcal{R}_k}
      \bigg|_{\mathcal{R}_k=\sqrt{\mathcal{B}_k^2+\mathcal{C}_k^2}}
    = \frac{f_\lambda}{\lambda} , \qquad
  I^{(2,0)}_k = \frac{1}{\mathcal{A}_k^2-\mathcal{B}_k^2-\mathcal{C}_k^2}
    = \frac{1}{sq^2} ,
\end{equation}
and all higher $(i,j)$ follow from two-term recursions in $j$
(Appendix~\ref{app:coefficients}), which are
applied in $d=4-2\epsilon$ where required: the classes whose
integrand has a collinear pole are evaluated through
$\order{\epsilon}$, because their $\order{\epsilon}$ terms multiply
the $1/\epsilon$ of the $s_2$ expansion of Eq.~\eqref{eq:s2exp} and
feed the finite plus-distribution terms. In our derivation these
integrals are organised in an independently constructed basis: the
single-denominator elements of Eq.~\eqref{eq:angularbasis} are the
$I^{(i,j)}_k$ themselves, and the multi-denominator ones are reduced
to them by exact partial fractioning in the two scalar products of
the emitted parton with the beams, whose polar
integrals close on three one-dimensional master families. After
this reduction every remaining polar integrand has the form
\begin{equation}
  g(x) = \frac{P_A(x) + P_B(x)/\sqrt{Q(x)} + P_C(x)/Q(x)^{3/2}}{x^{\,j}} ,
  \qquad j\le 2 ,
\end{equation}
where $x = 1\mp \cos\theta_1$ is the distance from the single endpoint pole
left by the outer denominators, the $P_\bullet$ are polynomials and
$Q(x) = \alpha_D^2-\beta_D^2 > 0$ is the discriminant of the single
inner azimuthal denominator $D = \alpha_D + \beta_D\cos\theta_2$
that survives the cascade (the
tangential denominators give perfect squares and are absorbed in
closed form). The outer integrals therefore reduce to the finite
parts (in the plain-cutoff sense used by the endpoint subtraction)
of
\begin{equation}
  R_m = \mathrm{FP}\!\int_0^2\! x^m\,\mathrm{d}x , \qquad
  J_m = \mathrm{FP}\!\int_0^2\! \frac{x^m}{\sqrt{Q(x)}}\,\mathrm{d}x , \qquad
  V_m = \mathrm{FP}\!\int_0^2\! \frac{x^m}{Q(x)^{3/2}}\,\mathrm{d}x ,
\end{equation}
with $R_{-1} = \ln 2$, $R_m = 2^{m+1}/(m+1)$ otherwise, the $J_m$
given by a square-root--logarithm recursion and the $V_m$ reduced
algebraically to the $J_m$; every such integral was gated
numerically against the quadrature evaluation of the same element.

\subsection{Virtual master integrals}
\label{app:notation:virtual}

The one-loop virtual corrections are reduced by
integration-by-parts identities to the standard one-loop basis for
one off-shell external leg. Seven masters remain: the massless
bubbles in $s$, $t$, $u$ and $q^2$, and the three one-mass boxes with
invariant pairs $(s,t)$, $(s,u)$ and $(t,u)$; the one-mass triangles
reduce to bubbles. In $d = 4-2\epsilon$ dimensions, with the
normalisation of Ref.~\cite{Ellis:2007qk} (renormalisation scale set to
one, invariants with a positive imaginary part), the bubble and the
one-mass box are
\begin{align}
  I_2(x) &= \frac{1}{r_\Gamma}\int\!\frac{\mathrm{d}^dk}{i\pi^{d/2}}\,
    \frac{1}{k^2\,(k+p)^2}
  = \frac{1}{\epsilon} + 2 - \ln(-x) + \order{\epsilon} ,
  \qquad p^2 = x , \notag\\
  I_4(x,y) &= \frac{1}{r_\Gamma}\int\!\frac{\mathrm{d}^dk}{i\pi^{d/2}}\,
    \frac{1}{k^2\,(k+p_1)^2\,(k+p_1+p_2)^2\,(k+p_1+p_2+p_3)^2}
  \notag\\
  &= \frac{1}{xy}\biggl\{ \frac{2}{\epsilon^2}\Bigl[(-x)^{-\epsilon}
    + (-y)^{-\epsilon} - (-q^2)^{-\epsilon}\Bigr]
    - 2\,\mathrm{Li}_2\Bigl(1-\frac{q^2}{x}\Bigr)
    - 2\,\mathrm{Li}_2\Bigl(1-\frac{q^2}{y}\Bigr)
    - \ln^2\frac{x}{y} - \frac{\pi^2}{3} \biggr\} + \order{\epsilon} ,
\end{align}
with $p_1^2 = p_2^2 = p_3^2 = 0$, $(p_1+p_2+p_3)^2 = q^2$,
$x = (p_1+p_2)^2$, $y = (p_2+p_3)^2$ and
$r_\Gamma = \Gamma^2(1-\epsilon)\,\Gamma(1+\epsilon)/\Gamma(1-2\epsilon)$;
the seven masters are $I_2(s)$, $I_2(t)$, $I_2(u)$, $I_2(q^2)$,
$I_4(s,t)$, $I_4(s,u)$ and $I_4(t,u)$. The bubbles contribute the single logarithms
$\ell_x$, the boxes the dilogarithms $\mathrm{Li}_2(1-q^2/x)$ and the
squared logarithm differences that assemble into the functions
$f^{(1)}$ and $f^{(2)}$ of Appendix~\ref{app:coefficients}. The full
$\epsilon$-expansion of the reduced amplitude, with symbolic
coefficients in $(t,u,q^2,N_c)$, is generated by the same chain that
produces the real-emission coefficient functions and is the object whose poles were
verified against the universal structure above. For the T-odd sector
only the absorptive parts of these masters enter, the imaginary parts
of the $s$-channel logarithms and dilogarithms, which are finite and
are evaluated in four dimensions; the axial triangles use the
representation of Section~\ref{sec:triangles}.

\section{Coefficient functions and ancillary files}
\label{app:coefficients}
\begin{fleqn}[0pt]

The complete set of closed-form coefficient functions is provided as
ancillary files with the arXiv submission, listed in
Table~\ref{tab:anc}: one plain-text and one \sympy-importable file per
sector, in the notation of this appendix. A \texttt{README} gives the
conventions, defines every symbol that appears in the files, and
states how each file enters the helicity cross sections, including
the signed crossing relations that are not repeated as separate
files and the assembly of the endpoint sector from its soft and
virtual parts. The
atoms and angular master integrals in which the coefficient functions
are written are defined below.

\begin{table}[htbp]
  \centering
  \caption{The ancillary files, one per sector ($ab$ the channel,
    $\beta$ the projection, $ij$ the four-quark class of
    Appendix~\ref{app:notation:couplings}).}
  \label{tab:anc}
  \begin{tabular}{p{0.30\textwidth}p{0.62\textwidth}}
    \toprule
    file & content \\
    \midrule
    \texttt{notation} & atoms and angular master integrals $I^{(i,j)}_k$ \\
    \texttt{born} & Born structure functions $C^{\beta,(0)}_{ab}$ \\
    \texttt{virtual} & endpoint sector: soft-endpoint ratios
      $C^{\beta,\delta}_{ab}/C^{\beta,(0)}_{ab}$ and one-loop virtual
      columns $B^{\beta}_{ab}$ \\
    \texttt{dk} & parity-odd virtual factors $dK_3$, $dK_4$ \\
    \texttt{hard\_qqb}, \texttt{hard\_qg}, \texttt{hard\_gg} &
      hard coefficient functions $C^{\beta,\log}_{ab}$, $C^{\beta,1}_{ab}$, $C^{\beta,\mathrm{reg}}_{ab}$
      (\texttt{hard\_gg} with its collinear bridge) \\
    \texttt{fourquark} &
      four-quark classes $C^{\beta}_{q\bar q,ij}$, $C^{\beta}_{qq',ij}$ \\
    \texttt{odd\_hard}, \texttt{odd\_cc4q} &
      parity-odd coefficient functions $C^{O_1,\mathrm{reg}}$, $C^{O_2,\mathrm{reg}}$ \\
    \texttt{todd} & T-odd coefficient functions $C^{\{7,8,9\},\delta}_{q\bar q,\,qg}$ \\
    \texttt{triangle\_even}, \texttt{triangle\_todd}, \texttt{triangle\_dabt} &
      axial-triangle sector of Section~\ref{sec:triangles} \\
    \texttt{triangle\_todd\_eval.py} &
      evaluator of the absorptive triangle contributions (simplex
      moments and assembly) \\
    \bottomrule
  \end{tabular}
\end{table}

Kinematics, projectors, couplings and luminosities, the
plus-distribution prescriptions, and the master integrals are
defined in Appendix~\ref{app:notation}; here we only add the
shorthand atoms in which the coefficient functions are written:
the logarithmic atoms $f_a$, the rational atoms $d_a$, the Gram
combination $\lambda$, and the angular integrals $I^{(i,j)}_k$ of
Eq.~\eqref{eq:angularclass} evaluated at the two propagator
families $k = 1, 2$ of the process, the propagators of the emitted
parton with the two beams.

The logarithmic and rational atoms, at general $s_2$ (they reduce to the $\delta(s_2)$-kinematics forms at $s_2=0$):
\begin{align}
& d_t = \frac{1}{s_2 - t} , \qquad d_u = \frac{1}{s_2 - u} , \qquad d_s = \frac{1}{q^2 + s - s_2} , \qquad d_{st} = \frac{1}{s - s_2 + t} , \qquad d_{su} = \frac{1}{s - s_2 + u} , \\
& d_{tu} = \frac{1}{- q^2 s_2 + t u} , \qquad \lambda = \sqrt{- 4 q^2 s_2 + (t + u)^{2}} , \qquad f_{s_2} = \ln\bigl(\frac{s_2}{q^2}\bigr) , \\
& f_{stu} = \ln\bigl(\frac{q^2 s}{(s_2 - t) (s_2 - u)}\bigr) , \qquad f_{tu} = \ln\bigl(\frac{- q^2 s_2 + t u}{(s_2 - t) (s_2 - u)}\bigr) , \qquad f_{st} = \ln\bigl(\frac{s t^{2}}{q^2 (s_2 - t)^{2}}\bigr) , \\
& f_{su} = \ln\bigl(\frac{s u^{2}}{q^2 (s_2 - u)^{2}}\bigr) , \qquad f_{\lambda t} = \ln\Bigl(\frac{q^2 s (s_2 - t)^{2}}{\bigl(- q^2 t + s_2 (2 q^2 - u)\bigr)^{2}}\Bigr) , \\
& f_{\lambda u} = \ln\Bigl(\frac{q^2 s (s_2 - u)^{2}}{\bigl(- q^2 u + s_2 (2 q^2 - t)\bigr)^{2}}\Bigr) , \qquad f_\lambda = \ln\bigl(\frac{q^2 + s - s_2 + \sqrt{- 4 q^2 s_2 + (t + u)^{2}}}{q^2 + s - s_2 - \sqrt{- 4 q^2 s_2 + (t + u)^{2}}}\bigr) , \\
& f_s = \ln\bigl(\frac{s}{q^2}\bigr) , \qquad f_t = \ln\bigl(- \frac{t}{q^2}\bigr) , \qquad f_u = \ln\bigl(- \frac{u}{q^2}\bigr) , \qquad f_t^{(1)} = \frac{\ln^{2}\bigl(\frac{q^2}{q^2 - t}\bigr)}{2} + \operatorname{Li}_{2}\bigl(\frac{q^2}{q^2 - t}\bigr) , \\
& f_u^{(1)} = \frac{\ln^{2}\bigl(\frac{q^2}{q^2 - u}\bigr)}{2} + \operatorname{Li}_{2}\bigl(\frac{q^2}{q^2 - u}\bigr) , \\
& f_t^{(2)} = \frac{\ln^{2}\bigl(\frac{s}{q^2}\bigr)}{2} + \ln\bigl(\frac{s}{q^2}\bigr) \ln\bigl(- \frac{t}{- q^2 + s}\bigr) + \operatorname{Li}_{2}\bigl(\frac{q^2}{s}\bigr) , \\
& f_u^{(2)} = \frac{\ln^{2}\bigl(\frac{s}{q^2}\bigr)}{2} + \ln\bigl(\frac{s}{q^2}\bigr) \ln\bigl(- \frac{u}{- q^2 + s}\bigr) + \operatorname{Li}_{2}\bigl(\frac{q^2}{s}\bigr) .
\end{align}

The endpoint ($\delta(s_2)$) coefficient functions carry in addition the two
scale logarithms $f_{M^2} = \ln(\muF^2/q^2)$ and
$f_A = \ln(A/q^2)$, where $A$ is the upper limit of the $s_2$ slice
that defines the $[\,\cdot\,]_{A+}$ distributions of
Appendix~\ref{app:notation:phasespace}; the $A$ dependence cancels
between the endpoint coefficient functions and the regular coefficient functions integrated
over $0<s_2<A$.

The constants $(\mathcal{A}_k, \mathcal{B}_k, \mathcal{C}_k)$ of
Eq.~\eqref{eq:angulardenom} for the two families, the base cases and
the recursion in $j$ of the angular integrals are, with
$s\,\qT^2 = tu - s_2 q^2$ ($\mathcal{A}_1 = \mathcal{A}_2$ for this
process, so the family label is dropped on $\mathcal{A}$, while
$\mathcal{R}_k = \sqrt{\mathcal{B}_k^2+\mathcal{C}_k^2} = \lambda/2$
holds exactly for both families):

\begin{align}
& \mathcal{A}_1 = \mathcal{A}_2 = q^2 - \frac{t}{2} - \frac{u}{2} , \qquad \mathcal{R}_k = \sqrt{\mathcal{B}_k^2+\mathcal{C}_k^2} = \frac{\lambda}{2} , \notag\\
& \mathcal{B}_1 = - d_u s s_2 + s_2 - \frac{t}{2} - \frac{u}{2} , \qquad \mathcal{C}_1 = d_u \qT s \sqrt{s_2} , \notag\\
& \mathcal{B}_2 = - d_t s s_2 + s_2 - \frac{t}{2} - \frac{u}{2} , \qquad \mathcal{C}_2 = d_t \qT s \sqrt{s_2} ,
\end{align}
\begin{equation}
  I^{(1,0)}_k = \frac{1}{2\mathcal{R}_k}\ln\frac{\mathcal{A}+\mathcal{R}_k}{\mathcal{A}-\mathcal{R}_k} = \frac{f_\lambda}{\lambda} , \qquad I^{(2,0)}_k = \frac{1}{\mathcal{A}^{2} - \mathcal{R}_k^{2}} = \frac{1}{q^2 s} ,
\end{equation}
\begin{align}
& I^{(1,j)}_k = \frac{1}{j\,\mathcal{R}_k^2}\Bigl[ \tfrac12\bigl[\mathcal{A} - \mathcal{B}_k + (-1)^j\,(\mathcal{A} + \mathcal{B}_k)\bigr] + (2j-1)\,\mathcal{A}\,\mathcal{B}_k\,I^{(1,j-1)}_k - (j-1)\,(\mathcal{A}^{2} - \mathcal{C}_k^{2})\,I^{(1,j-2)}_k \Bigr] , \notag\\
& I^{(2,j)}_k = \frac{1}{j\,\mathcal{R}_k^2}\Bigl[ -\tfrac12\bigl[1+(-1)^j\bigr] - (2j-1)\,\mathcal{B}_k\bigl(I^{(1,j-1)}_k - \mathcal{A}\,I^{(2,j-1)}_k\bigr) \notag\\
& \qquad\qquad + (j-1)\bigl[2\mathcal{A}\,I^{(1,j-2)}_k - (\mathcal{A}^{2} - \mathcal{C}_k^{2})\,I^{(2,j-2)}_k\bigr] \Bigr] ,
\end{align}
\begin{equation}
  I^{(1,1)}_k = \frac{\mathcal{B}_k \bigl(\mathcal{A} I^{(1,0)}_{k} - 1\bigr)}{\mathcal{R}_k^{2}} , \qquad I^{(2,1)}_k = \frac{\mathcal{B}_k \bigl(\mathcal{A} I^{(2,0)}_{k} - I^{(1,0)}_{k}\bigr)}{\mathcal{R}_k^{2}} ,
\end{equation}

\end{fleqn}

\section{Misprints found in the literature}
\label{app:misprints}

The $\order{\as^2}$ coefficient functions of the $U{+}L$, $L$ and
transverse projections were computed in Ref.~\cite{Mirkes:1992hu},
whose T-odd appendix transcribes Ref.~\cite{Hagiwara:1984hi}. Of the
four-quark classes, Ref.~\cite{Mirkes:1992hu} prints the
interference classes $D_{ac}$ and $D_{ad}$; the classes $D_{ab}$,
$D_{cd}$, $E_{ab}$ and $E_{cd}$, which
contribute to neutral-current production only, are not printed
there but in the doctoral thesis of Ref.~\cite{Mirkes:1990diss}, to
which Ref.~\cite{Mirkes:1992hu} defers them; the vector--vector half
of $D_{cd}$ was later reproduced in the appendix of
Ref.~\cite{Mirkes:1994dp}. By comparing our coefficient functions to
these previous results we found a number of misprints, which we
record here since to our knowledge no erratum exists for any of
them. In this appendix the coefficient functions are named as in
Refs.~\cite{Mirkes:1992hu,Mirkes:1990diss}, whose common notation is
bridged to ours in Table~\ref{tab:bridge}. The comparison is symbolic: the printed expression, after
the repairs listed below where these are needed, is reduced to our
basis of logarithmic and rational atoms, and the difference with
our coefficient function vanishes coefficient by coefficient as a rational
function of the invariants, with $N_c$ a free symbol; each such
comparison carries a control in which a single printed coefficient
is perturbed and the difference is required not to vanish. In this
way every previously available coefficient function that we use,
including the repairs of the items below (and a few obvious
typographical readings of subscripts and parentheses, not listed),
was verified symbolically. This covers the four-quark classes of
Ref.~\cite{Mirkes:1990diss} as well: although they are absent from
the published literature, they provide a complete independent
analytic reference for the axial--axial $D_{cd}$ and the $D_{ab}$
classes. The coefficient functions
that have no counterpart anywhere in the literature and appear here
for the first time --- the parity-odd $\order{\as^2}$ coefficient
functions, including their charged-current four-quark classes, and
the parity-odd virtual factors $dK_i$ --- are verified against
numerical benchmarks only, as are the repairs below
(\textsc{OpenLoops}, \mcfm).

\begin{table}[htbp]
  \centering
  \caption{Correspondence between the coefficient functions of
    Appendix~\ref{app:coefficients}, in the notation of
    Eq.~\eqref{eq:s2structure}, and the objects of
    Refs.~\cite{Mirkes:1992hu,Mirkes:1990diss} ($\beta$ the projection,
    $ab$ the channel, $ij$ the four-quark class).}
  \label{tab:bridge}
  \begin{tabular}{p{0.40\textwidth}p{0.54\textwidth}}
    \toprule
    this paper & Refs.~\cite{Mirkes:1992hu,Mirkes:1990diss} \\
    \midrule
    $C^{\beta,(0)}_{ab}$ & Born structure functions $A^\beta_{ab}$ \\
    $C^{\beta,\delta}_{ab}$ & endpoint coefficients $C^\beta_{1,ab}$ (containing the one-loop virtual $B^\beta_{ab}$) \\
    $C^{\beta,\log}_{ab}$, $C^{\beta,1}_{ab}$, $C^{\beta,\mathrm{reg}}_{ab}$ &
      the $[\ln s_2/s_2]_{A+}$, $[1/s_2]_{A+}$ and regular parts of $C^\beta_{2,ab}$ \\
    $C^{\beta,\cdot}_{q\bar q,ij}$, $C^{\beta,\cdot}_{qq',ij}$ & four-quark classes $D^\beta_{ij}$, $E^\beta_{ij}$,
      with our attachment labels $(i,p,1,2)$ for their $(a,b,c,d)$ \\
    $C^{O_{1,2},\mathrm{reg}}_{ab}$ & parity-odd hard coefficient functions (not published) \\
    $C^{\{7,8,9\},\delta}_{ab}$ & T-odd structure functions $T^{7,8,9}$ (Appendix B of Ref.~\cite{Mirkes:1992hu}) \\
    $I^{(i,j)}_k$ of Eq.~\eqref{eq:angularclass} & angular integrals $H_k^{(i,j)}$ (Appendix F of Ref.~\cite{Mirkes:1992hu}) \\
    \bottomrule
  \end{tabular}
\end{table}

The following misprints were found in Ref.~\cite{Mirkes:1992hu}:
\begin{enumerate}
\item \textbf{$qg$ hard coefficient function, $q_7$ term.} A term is
  dropped; the correct form is
  $q_7 = -4t\, d_{st}\,(s+s_2-q^2)$.
\item \textbf{$B^{L12}_{q\bar q}$ (virtual).} The bracket
  $+\frac{N_c}{2}\left[ f_t\!\left(\frac{t}{s+u}-2\right) +
  f_u\!\left(\frac{u}{s+t}-2\right)\right]$ has the wrong overall
  sign, and the constant $-(C_F-\frac{N_c}{2})\frac{\pi^2}{6}$ is
  spurious.
\item \textbf{$B^{L2}_{qg}$ and $B^{L12}_{qg}$ (virtual).} Both
  braces are missing $+(C_F-\frac{N_c}{2})\frac{\pi^2}{6}$ --- the
  same constant that item 2 spuriously adds to the $q\bar q$
  coefficient function, suggesting a transposition during preparation. In
  addition, the $1/\epsilon$ bracket of $B^{L2}_{qg}$ is printed as
  $N_c(f_s-f_u-f_t)$, whereas the other three $qg$ projections print
  $N_c(f_u-f_s-f_t)$; since the infrared poles multiply the Born
  tensor, the bracket is projection independent and the latter form
  is the correct one.
\item \textbf{$C^{L1}_{2,q\bar q}$ hard coefficient function (p.~40).} One sign:
  the last brace must read
  \begin{equation*}
    + C_F d_t\left(\frac{s_2-u}{s}
      - s_2\left(d_t - \frac{2}{u}\right)\right)
  \end{equation*}
  (printed with $+s_2(\ldots)$).
\item \textbf{$C^{L2}_{2,qg}$ hard coefficient function.} Two single-token
  misprints, one in each colour structure of
  $C^{L2}_{2,qg} = N_c A^{L2}_{qg} + C_F B^{L2}_{qg}$. In
  $A^{L2}_{qg}$, inside the $-d_t d_u/(q^2 s t^4)$ brace, the group $s(s_2-t)[\ldots]$
  contains $8u^2 t s_2\,(31 s_2 t + 12 u s_2 - 8ut)$, whose last
  coefficient must read $-7ut$. In $B^{L2}_{qg}$, inside the
  $-1/(16 s)$ brace, the $H_2^{(1,1)}$ term
  $4(s_2-t)\,d_{st}\,[\ldots]$ is missing an overall factor $1/t$;
  the printed $(s_2-t)$ is genuine. The $qg$ coefficient
  functions are set a second time, independently, in
  Ref.~\cite{Mirkes:1990diss}: there the first of the two terms
  reads $-7ut$, so that defect entered in the preparation of
  Ref.~\cite{Mirkes:1992hu}, whereas the missing $1/t$ of the
  second is printed the same way in both.
\item \textbf{$C^{\beta}_{2,gg}$ hard coefficient functions.} The printed $gg$
  coefficient functions are $N_c A_{gg} + C_F B_{gg}$, and the two colour
  structures carry independent defects, which only separate once the
  comparison is graded by colour. Seven misprints are identified.
  (i) In $B_{gg}$, in all three coefficient functions, the bracket multiplying the
  factorisation-scale logarithm is short by one unit: it must read
  $(f_{M^2} - f_{s_2} + \ln 2 + 1)$, the printed bracket carrying no
  $+1$ (the same unit may equivalently be read as a $-1$ beside
  $f_{s_2}$ or a $+1$ beside $\ln 2$; the difference alone cannot
  discriminate, and $A_{gg}$ has no factorisation-scale bracket at
  all). (ii) In $A^{L1}_{gg}$ the $f$-block $(f_{\lambda u}-f_{su})$
  must read $(f_{\lambda u}+f_{su})$: the required correction is
  exactly $-2$ times the printed coefficient, the signature of a
  sign flip. (iii) In $B^{L1}_{gg}$ the $f_{su}$ bracket carries
  $4t$ where the correct coefficient is $3t$. (iv) In $A^{L12}_{gg}$
  the $f_{\lambda u}$ and $f_{\lambda t}$ brackets are printed with
  an asymmetric pair of coefficients: the printed
  $3(2u^2+4s_2^2+ut)$ of the $f_{\lambda u}$ member must read
  $3(2u^2+3s_2^2+ut)$, the
  $t\leftrightarrow u$ image of its partner. (v) In $B^{L12}_{gg}$,
  one coefficient of an $H^{(1,0)}_1$ bracket: the printed
  $-2s^2(34s_2^2 + 28tu + 19u(u-4s_2))$ must read $26tu$; the two
  printed occurrences of $H^{(1,0)}$ are $t\leftrightarrow u$ images
  of the same function, so the comparison cannot say which of them
  carries the slip. (vi) In $A^{L1}_{gg}$, the first term of the
  $H_1^{(1,2)}$ bracket is printed as $s_2(14s+21u+25t-36s_2)$ and
  must read $s^2(14s+21u+25t-36s_2)$: a superscript typeset as a
  subscript. (vii) In $A^{L1}_{gg}$, the $H_2^{(1,2)}$ bracket
  $t(5t-11s+3u)$ must read $t(5t+11s+3u)$, a sign flip. The last two
  sit in different terms of the same colour structure and add up in
  a way that no single-token search can resolve. Of the seven,
  (i), (iv) and (v) belong to $L12$ and (i)--(iii), (vi) and (vii)
  to $L1$ and $L2$, which are $t\leftrightarrow u$ images of each
  other.
\item \textbf{$D^{L12}_{ac}$ (p.~43).} A binary $+$ is swallowed
  at a line break inside the $H_2^{(1,1)}$ bracket, between the
  fragments $\bigl(u(u+t-2s_2)+t(t-s_2)\bigr)(u+t-2s_2)$ and
  $s(u+3t-6s_2)(u+t)$, which the print sets on consecutive lines
  with no operator between them; both are of degree three, so what
  is lost is an operator and not a factor. $D^{L12}_{ad}$ follows by
  $t\leftrightarrow u$ crossing.
\item \textbf{$g\to q\bar q$ ($n_f$) $L1/L2$ regular coefficient functions.}
  A dropped factor $s/2$ renders the printed terms
  dimensionally inhomogeneous:
  $(s_2-2u)/u^2 \to s\,(s_2-2u)/(2u^2)$ and
  $t\leftrightarrow u$.
\item \textbf{Angular constants $B_k$ of (F.9).} The printed
  $B_1$/$B_2$ are swapped between the two axis families (the $C_k$
  are correct): within one family the $\cos\vartheta$ coefficient
  of $B_k$ and the $\sin\vartheta$ coefficient of $C_k$ must carry
  the same scale. The same swap is printed in
  Ref.~\cite{Mirkes:1990diss}, so it precedes
  Ref.~\cite{Mirkes:1992hu}.
\item \textbf{$D^{L1}_{ac}$ superscripts.} The printed
  $H_1^{(0,1)}$ is $H_1^{(1,0)}$ (a transposed superscript;
  $(0,1)$ vanishes identically), and the printed $H_3^{(1,1)}$,
  $H_3^{(1,2)}$ are $H_2^{(1,1)}$, $H_2^{(1,2)}$ (no $k=3$ family
  exists).
\end{enumerate}
After these repairs each printed coefficient function of
Ref.~\cite{Mirkes:1992hu} that we use is symbolically equivalent to
the one derived here, for general $N_c$.

The following misprints were found in Appendix~B of
Ref.~\cite{Mirkes:1992hu} and in Ref.~\cite{Hagiwara:1984hi}.
Appendix B of Ref.~\cite{Mirkes:1992hu} is a variable-converted
transcription of the T-odd structure functions of
Ref.~\cite{Hagiwara:1984hi}. We find three defects introduced in
the transcription and one in the source; the factors below were
confirmed against the original functions of
Ref.~\cite{Hagiwara:1984hi}:
\begin{enumerate}
\setcounter{enumi}{10}
\item $T^8$ is over-normalised by $2\sqrt{2}$ in both channels,
  relative to the normalisation of Eq.~(15) of
  Ref.~\cite{Mirkes:1992hu};
\item $T^9$ is over-normalised by $8$ in the same sense, so that
  $T^7$, $T^8$ and $T^9$ are printed $1$, $2\sqrt2$ and $8$ times the
  functions normalised as in Eq.~(15) of Ref.~\cite{Mirkes:1992hu}
  (these factors are not the normalisations of $A_5$, $A_6$, $A_7$ in
  Appendix~\ref{app:notation:projectors});
\item in the $qg$ block of the appendix the signs of $T^8_{qg}$ and
  $T^9_{qg}$ are reversed with respect to its own $q\bar q$ block and
  to Ref.~\cite{Hagiwara:1984hi}, whose two channels are mutually
  consistent;
\item in Ref.~\cite{Hagiwara:1984hi} itself, the $C_1$ bracket of
  the function $f_{A_9}$ in Eq.~(8) is printed with a minus sign
  and must read $\left[1 + \frac{c}{c-a}\ln\frac{a}{c}\right]$.
  This is, we believe, exactly the misprint alluded to in the
  footnote of Ref.~\cite{Mirkes:1992hu}. In
  Ref.~\cite{Hagiwara:1984hi} the functions $F_7$, $F_8$ and $F_9$
  multiply $\sin\theta\sin\phi$, $\sin2\theta\sin\phi$ and
  $\sin^2\theta\sin2\phi$, i.e.\ they correspond to $A_7$, $A_6$ and
  $A_5$ of Eq.~\eqref{eq:angulardecomposition}, in reverse order;
  $f_{A_9}$ therefore feeds $A_5$, the largest T-odd coefficient. With this single
  sign repaired, the 1984 result becomes a complete independent
  analytic benchmark for the T-odd sector, agreeing with our
  derivation in both channels up to the per-projection normalisation
  constants above.
\end{enumerate}

The following misprints were found in the four-quark classes of
Ref.~\cite{Mirkes:1990diss}, which prints $D_{ab}$ --- whose
vector--vector part vanishes by Furry's theorem --- and both the
vector--vector and the axial--axial halves of $D_{cd}$, in all four
projections, and gives $E_{ab}$ and $E_{cd}$ as crossings of
$D_{cd}$. Items 15--17 below sit in the vector--vector half of
$D_{cd}$ and are also printed, verbatim, in the appendix of
Ref.~\cite{Mirkes:1994dp}, which was typeset from it; item 18
affects only the half that was never published:
\begin{enumerate}
\setcounter{enumi}{14}
\item \textbf{$D^{U+L}_{cd,VV}$}: three $d$-atom substitutions in
  the logarithmic block ($d_u d_t \to d_{st} d_{su}$; $d_s d_u \to
  d_s d_{st}$; $d_{st} d_s \to d_{su} d_s$).
\item \textbf{$D^{L1}_{cd,VV}$} (hence $L2$): the prefactor of the
  $H$-function brace must read $-d_{su}/(8u)$ (printed $+$).
\item \textbf{$D^{L12}_{cd,VV}$}: in the $+d_{st}/(2ut)\{\cdots\}$
  brace the printed $f_{\lambda t}\,d_{su}\,[\cdots]$ must read
  $f_{\lambda t}\,d_{s}\,u\,[\cdots]$ --- the factor $u$ multiplying
  the bracket has been absorbed into the subscript of $d_s$. The
  printed form is dimensionally inhomogeneous, which is how the
  misprint is first seen, but homogeneity does not fix it: deleting
  $d_{su}$ restores it too, and only the exact difference selects
  $d_s u$.
\item \textbf{$D^{L1}_{cd,AA}$, $D^{L2}_{cd,AA}$ and
  $D^{L12}_{cd,AA}$: overall normalisation.} The three axial--axial
  $L$ coefficient functions are printed at one half of the
  normalisation carried by the four vector--vector projections and
  by $D^{U+L}_{cd,AA}$, which share one class constant; the latter
  normalisation is the one that reproduces our result and the
  numerical benchmark. Since this is an overall factor it cannot be
  attributed to any particular symbol of the print; the two halves
  agree term by term on the part they share.
\end{enumerate}
After these repairs each four-quark coefficient function printed in
Ref.~\cite{Mirkes:1990diss} is likewise symbolically equivalent to
the one derived here: on the atom basis
$\{1, f_\lambda, f_{tu}, f_{st}, f_{su}, f_{\lambda t}, f_{\lambda
  u}\}$ every coefficient of the difference vanishes as a rational
function of $(t,u,s_2)$, with the single class constant
$4C_F/N_c$ ($=16/9$ at $N_c=3$) read off independently from each
atom--coefficient function pair. The four projections of $D_{ab}$
and $D^{U+L}_{cd,AA}$ are reproduced verbatim, with no repair at
all, and the axial--axial $L$ projections need none beyond the
overall factor of item 18. The printed four-quark classes thus
provide a complete independent analytic check of this sector, which
the published material alone could not: it contains one of the two
halves of one of the two classes.

\clearpage
\printbibliography

\end{document}